\documentclass[12pt]{article}
\usepackage{putex}
\usepackage{graphicx}

\begin{document}

\def \be{\begin{equation}}
\def \ee{\end{equation}}
\def \bd{\begin{displaymath}}
\def\ed{\end{displaymath}}
\def\dim{\mathop{dim}}
\def\NLsM{NL$\sigma$M}

\preprint{hep-th/0605102 \\ PUPT-2197}

\institution{PU}{Joseph Henry Laboratories, Princeton University, Princeton, NJ 08544}

\title{Heterotic non-linear sigma models with anti-de Sitter target spaces}

\authors{Georgios Michalogiorgakis and Steven S. Gubser}

\abstract{We calculate the beta function of non-linear sigma models with $S^{D+1}$ and $AdS_{D+1}$ target spaces in a $1/D$ expansion up to order $1/D^2$ and to all orders in $\alpha'$.  This beta function encodes partial information about the spacetime effective action for the heterotic string to all orders in $\alpha'$.  We argue that a zero of the beta function, corresponding to a worldsheet CFT with $AdS_{D+1}$ target space, arises from competition between the one-loop and higher-loop terms, similarly to the bosonic and supersymmetric cases studied previously in \cite{fg}.  Various critical exponents of the non-linear sigma model are calculated, and checks of the calculation are presented.}

\date{May 2006}

\maketitle

\tableofcontents

\section{Introduction}
\label{INTRODUCTION}

Particular interest attaches to backgrounds of string theory involving $AdS_{D+1}$
because of their relation to conformal field theories in $D$ dimensions
\cite{juanAdS,gkPol,witHolOne} (for a review see \cite{MAGOO}).  But because these
geometries (with some exceptions) arise from the near-horizon geometry of D-branes,
formulating a closed string description is complicated by the presence of Ramond-Ramond fields.

 It was recently proposed \cite{fg} that $AdS_{D+1}$ vacua might
exist without any matter
 fields at all.  Instead of relying upon the stress-energy of matter fields to curve space,
 the proposal is that higher powers of the curvature compete with the Einstein-Hilbert term
 to produce string-scale $AdS_{D+1}$ backgrounds.  The main support for this proposal comes
 from large $D$ computations of the beta function for the quantum field theory on the
 string worldsheet.  Before discussing these computations, let us review the lowest-order corrections to the beta function in an $\alpha'$ expansion:
 \eqn{LeadingCorrection}{\seqalign{\span\TT\qquad & \span\TR}{
  bosonic: & \beta_{ij} = \alpha' R_{ij} + {\alpha'^2 \over 2}
    R_{iklm} R_j{}^{klm} + O(\alpha'^3)  \cr
  heterotic: & \beta_{ij} = \alpha' R_{ij} + {\alpha'^2 \over 4}
    R_{iklm} R_j{}^{klm} + O(\alpha'^3) \cr
  \hbox{type~II:} & \beta_{ij} = \alpha' R_{ij} + \frac{\zeta(3)
   \alpha'^4}{2} R_{mhki}R_{jrt}{}^m (R^k{}_{qs}{}^r R^{tqsh} +
     R^k{}_{qs}{}^t R^{hrsq}) + O(\alpha'^5) \,.
 }}
These expressions are obtained using dimensional regularization with minimal subtraction,
and all derivatives of curvature are assumed to vanish as well as all matter fields.
Derivatives of curvature indeed vanish for symmetric spaces: for example,
 \eqn{AdScurvature}{
  R_{ijkl} = -{1 \over L^2} (g_{ik} g_{jl} - g_{il} g_{jk})
 }
in the case of $AdS_{D+1}$.  One indeed finds non-trivial zeroes for $AdS_{D+1}$ from all
 three beta functions in \eno{LeadingCorrection}.  An examination of higher order
 corrections in the bosonic and type~II cases shows that the zero persists in the most
  accurate expressions for the beta function that are available at present; however its
  location changes significantly, converging to $\alpha' D/L^2 = 1$ as $D$ becomes large.
   One aim of the present paper is to pursue similar large $D$ computations in the heterotic case.

 It should be clear from the outset that the question of the
existence of $AdS_{D+1}$ vacua with $\alpha' D/L^2$ close to unity
is a difficult one to settle perturbatively. Fixed order
computations are not reliable guides because the scale of curvature
is close to the string scale.  Large $D$ computations with finite
$\alpha' D/L^2$ seem to be a better
 guide, but they too could be misleading, mainly because higher order effects in $1/D$ than
 we are able to compute could change the behavior of the beta function significantly.
 These difficulties were discussed at some length in \cite{fg}.  Also, the vanishing of a
 beta function such as the ones in \eno{LeadingCorrection} is only a necessary condition for
 constructing a string theory: one must also cancel the Weyl anomaly and formulate a GSO
  projection that ensures modular invariance and the stability of the vacuum.

 There is a more general reason to be interested in high-order
computations of the beta function on symmetric spaces: from them we
can extract information about the structure of high powers of the
curvature that is quite different from what is available from
expansions of the Virasoro amplitude. While the latter tells us
about terms involving many derivatives but only four powers
 of the curvature (because only four gravitons are involved in the collision), the former tells
  us about many powers of the curvature with no extra derivatives.

The organization of the rest of this paper is as follows.  In section~\ref{GENERALFEATURES}, some general properties
 of the heterotic \NLsM\ are discussed.  In section~\ref{EXPONENTS}, the formalism and the results at $1/D$ order
  are presented.  In section \ref{NEXTORDER}, the critical exponents at $1/D^{2}$, the beta function,
  and the central charge of the CFT are computed.  The appendices include a brief explanation
  of the method of the calculation for the diagrams needed and the values of these diagrams.

\section{The heterotic non-linear sigma model}
\label{GENERALFEATURES}

As in \cite{fg}, much will be made of a connection through analytic continuation of the
NL$\sigma$M on $AdS_{D+1}$ and the NL$\sigma$M on $S^{D+1}$.  If $L$ is the radius of
$S^{D+1}$ and $g = \alpha'/L^2$, then continuing to negative $g$ leads to the $AdS_{D+1}$
 NL$\sigma$M.  The argument in \cite{fg} is slightly formal because it relies on an
 order-by-order perturbative evaluation of the partition function.

\begin{figure}[t]
  \centering
  \includegraphics[scale=0.8]{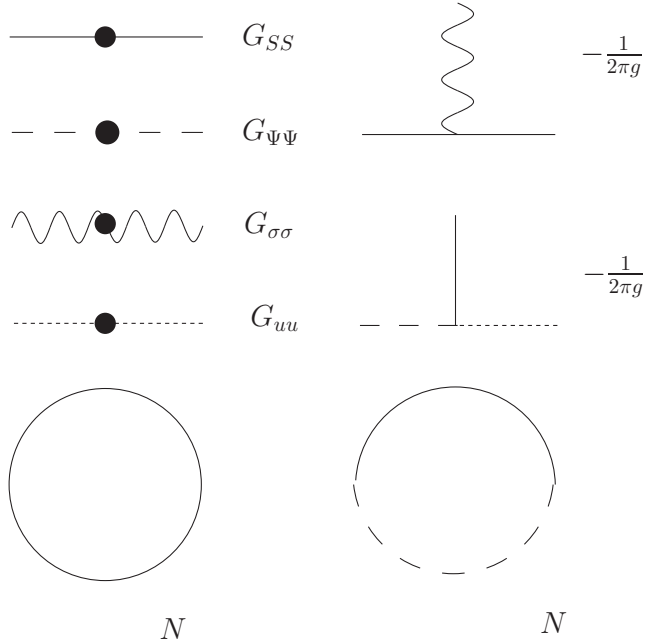}
  \caption{The Feynman rules for the heterotic sigma model.  The shaded circles indicate a dressed propagator.
  The circles indicate that a loop involving only the components of the $\Phi$ superfield receives a factor of $N$.
  We have suppressed the tensor structure
of the rules since only $\delta_{\mu \nu}$ appears.}\label{frules}
 \end{figure}

The action for the $S^{D+1}$ heterotic NL$\sigma$M is
 \begin{equation}\label{hetaction}
 S = \frac{1}{4\pi g} \int d^{2} x  d \bar{\theta} \left[ D_{+} \Phi
 \partial_{-} \Phi + \Lambda( \Phi^{2} -1)\right] + \frac{1}{4\pi g}
 \int d^{2} x \lambda_{A}\partial_{+}\lambda_{A}
 \end{equation}
where
 \eqn{Sfields}{
  \Phi = S + \bar{\theta} \Psi \qquad
   \Lambda = u  + \bar{\theta} \sigma \qquad
   D_+ = \frac{\partial }{\partial \bar{\theta}} +
     \bar{\theta}\frac{\partial}{\partial x_{+}} \qquad
   \partial_\pm = \frac{\partial}{\partial x_\mp} \,.
 }
$\Lambda$ is a spinorial superfield, and $u$ and $\Psi$ have opposite chirality. This leads to the action
 \begin{equation}\label{action}
  S = \frac{1}{4\pi g} \int d^{2}x \left[ (\partial S)^{2} +
  \bar{\Psi}i\partial \Psi + \sigma ( S^{2}-1) + 2 \bar{u} \Psi
  S\right] \,.
 \end{equation}
We have omitted the fermions $\lambda_A$ from \eno{action} because they decouple from
the gravitational action when the gauge field is set to zero \cite{AS,WH} as in our case.
 The Feynman rules for the theory \eno{action} can be seen in figure~\ref{frules}.  There is
 also a tadpole for $\sigma$, but we omit it because it does not contribute to the Dyson equations for the
 scaling parts of the dressed propagators, as in \cite{Vone}.

After a change of variables that renders the kinetic terms canonical, we can continue
to negative values of $g$ as in \cite{fg} to obtain an $AdS_{D+1}$ heterotic \NLsM.
 Quantities that are computed locally and perturbatively, such as $n$-point functions, cannot
 distinguish between a space of positive or negative curvature. As the beta function is
 derived from such quantities, it too can be continued to negative $g$, at least order by order
 in perturbation theory.

The heterotic NL$\sigma$M on $S^{D+1}$ is a generalization of the $O(D+2)$
model, and much of the relevant literature concentrates on an
expansion in $1/(D+2)$ rather than $1/D$.  We will
 therefore set
 \eqn{NvsD}{
  N = D+2
 }
and work with $N$ or $D$, according to convenience, in the rest of this paper.

\subsection{Some properties of the heterotic \NLsM\ for large $D$}
\label{LARGED}

 It is known that in the bosonic sigma model a mass appears \cite{AP}
 in the 1/N expansion.  The same phenomenon appears  in the supersymmetric
 extension of the sigma model \cite{witSigma,OA} where also the fermions
 acquire the same mass, signaling chiral symmetry breaking.
  In the heterotic case the bosons $S$ also acquire the same mass,
 showing that the interaction term does not destroy this effect.  To
 understand this, let's start from our action \eno{action}, and in the partition function
 integrate first the fermionic fields and then the bosons, since the action is quadratic
 in these.  We have omitted normalization factors of the partition function in the following.
\eqn{partitionfunction}{
Z &= \int \mathcal{D}S \mathcal{D} \Psi
\mathcal{D}\sigma \mathcal{D}u \, \exp\left(-\frac{i}{4\pi g} \int
d^{2}x [ (\partial S)^{2} + i \bar{\Psi} \slashed{\partial} \Psi +
\sigma(S^{2}-1) + 2 \bar{u}( S \cdot \Psi)]\right) \cr
 &= \int
\mathcal{D}S  \mathcal{D}\sigma \mathcal{D}u \,
\big[\det(i\slashed{\partial})\big]^{N/4} \exp\left[ \frac{i}{4\pi g} \int
d^{2} x  \left( S(\partial^{2} -\sigma)S- S^{i} \bar{u}
\frac{1}{i\slashed{\partial}} u S^{i} \right) \right]\cr
 &=\int
\mathcal{D}\sigma \mathcal{D}u \,
\big[\det(i\slashed{\partial})\big]^{N/4}
\left[\det\left(-\partial^{2}- \sigma \bar{u}
\frac{1}{i\slashed{\partial}} \bar{u}\right)\right]^{-N/2} \exp\left(
\frac{i}{4\pi g} \int d^{2} x \, \sigma\right) \cr &   \Rightarrow Z = \int
\mathcal{D}\sigma \mathcal{D}u \, e^{i S_{\rm eff}} \,,
 }
  where the
effective action for the Lagrange multiplier fields is given by
\def\Tr{\mop{Tr}}
\eqn{effact}{
  S_{\rm eff}  =  \int d^{2} x \left[\frac{1}{4\pi g}
\sigma - \frac{N}{4} \Tr \log (i \slashed{\partial}) +  \frac{N}{2}
\Tr\log\left(-\partial^{2}-\sigma -
\bar{u}\frac{1}{i\slashed{\partial}}u\right)\right] \,.
 }
 Since we are taking the limit $N \rightarrow \infty$ with $g_0 N$ finite,
 we see that all terms in the action are of order $N$.  We can evaluate
 this integral by the method of steepest descent, i.e.~by finding the classical value of
 $\sigma$, $u$ that minimizes the exponent, as is done for instance in \cite{AKM,GN}.
  This gives the variational
 equations
 \eqn{variation}{
 \langle x|\frac{1}{-\partial^{2}-   \sigma(x)
 -\bar{u}\frac{1}{i\slashed{\partial}}u} |x \rangle& = \frac{1}{2\pi
 N g} \cr
\langle x|\frac{\frac{1}{\slashed{\partial}}u}{-\partial^{2}-
\sigma(x)
 -\bar{u}\frac{1}{i\slashed{\partial}}u}|x \rangle & = 0 \,.
}
Because the right hand sides are constant, the left hand sides must
also be constant.  A solution to these equations is given by
\eqn{solutvariation}{
 u(x) = 0 \quad  \quad \sigma(x) = -m^{2} \,.
  }
 It is easy to see that $ \frac{1}{\partial} u(x) = const.$ has as its
 only solution $ u=0$.  This is in contrast to the supersymmetric
 case \cite{AKM}, where there are three solutions.  Now $m^{2}$ must satisfy
\eqn{masssquared}{
 \int d^{2} k \frac{1}{k^{2}+m^{2}} =
\frac{1}{2\pi g_{0}} \,.
 }
 Using a
simple-momentum cutoff, $ \frac{1}{2\pi g_{0}N} = \frac{1}{2\pi}
\log\frac{\Lambda}{m}$. By renormalizing at a scale $M$ we get
\eqn{oneloop}{
 \frac{1}{2\pi g_{0} N} = \frac{1}{ 2\pi g N} +
\frac{1}{2\pi} \log\frac{\Lambda}{M}\,.
 }
 Solving for the mass $m$ we get
$ m = M \exp[-1/gN]$.  Since this is a physical mass we expect that
it does not depend on the renormalization scale.  Using the
Callan-Symanzik equation for $m$,
 \eqn{callansymanzik}{
\left( M\frac{\partial}{\partial M} + \beta(g)\frac{\partial}{\partial g} \right)
m(g,M) = 0 \,,
 }
 gives the beta function $ \beta(g) = - g^{2}N$.  The mass is the
same in the bosonic, supersymmetric, \cite{AP,OA} and heterotic
model. This could have been predicted since the first order $\beta $
function is the same in all models, $ \beta_{ij} = \alpha' R_{ij}$.
Another way to get the same result is to calculate the effective
potential for the $\sigma$ field and see that the minimum of the
potential is not at zero but at $ \sigma = M e^{-1/gN}$.  One can go
further and examine the effective action \eno{effact}.  It is easy
to evaluate the counterterms needed for one-loop renormalization, as
we have already computed the wave function renormalization of the
$\sigma$ field, and doing so one finds
 \eqn{effecticelangrangian}{
\mathcal{L}_{0,\rm eff} = \frac{1}{4\pi g} \left(1+gN \log
\frac{\Lambda^{2}}{M^{2}} \right)\sigma -  \frac{N}{4} \Tr
\log\slashed{\partial} + \frac{N}{2} \Tr \log\left(-\partial^{2} -\sigma -
\bar{u} \frac{1}{i\slashed{\partial}} u\right) \,.
  }
 The bare and the dressed
quantities are related by
  \eqn{baredressde}{
  \sigma_{0} =Z \sigma
 \qquad u_{0} =Z^{1/2} u \qquad g_{0} = Z^{-1} g\qquad Z =1 +
 \frac{gN}{2} \log \frac{\Lambda^{2}}{M^{2}} \,.
 }
 Calculating the quadratic terms in the fields $u$, $\sigma$ will give us
 the propagators for these fields.  We can easily find
 \eqn{effectiveaction}{
 S_{\rm eff} = \frac{N}{2} Tr
 \frac{1}{-\partial^{2}-m^{2}}
 \bar{u}\frac{1}{i\slashed{\partial}-m}u - \frac{N}{4} \Tr
 \frac{1}{-\partial^{2} -m^{2}} \sigma \frac{1}{-\partial^{2}-m^{2}}
 \sigma \,.
 }
 The next step is to evaluate the propagators.  One finds
 \cite{OA,AKM}
 \eqn{uprop}{
 S_{u}(k) = -\frac{2i}{N}(\slashed{k} -2m) V(k^{2}) \qquad D_{\sigma}(k)  =
 \frac{2i}{N}(4m^{2}-k^{2}) V(k^{2}) \,,
 }
 where in $d$ dimensions
 \eqn{prop}{
 V(k^{2}) =  \frac{(4 \pi)^{d/2}}{4\Gamma(2-d/2)} \left(
 \frac{4m^{2}-k^{2}}{4} \right)^{1-d/2}
 \left(_{2}F_{1}(2-d/2,1/2,3/2;\frac{k^{2}}{k^{2}-4m^{2}})\right)^{-1}
   \,.
 }
In two dimensions this simplifies to
 \def\arcone{\rm arctanh}
 \def\arctanh{\mop{arctanh}}
 \eqn{Vtwodim}{
 V(k^{2}) = \pi \sqrt{\frac{k^{2}}{k^{2}-4m^{2}}}\left( \arctanh \sqrt{\frac{k^{2}}{k^{2}-4m^{2}}}
 \right)^{-1} \,.
 }
It is easy to see that the $u$ propagator in $d=2$ dimensions has  a
pole at $ k^{2} =4m^{2}$.  This means that there is a particle
with this mass.  Since the classical equations give $ u = - i
S\slashed{\partial} \Psi$, there is a boson-fermion
bound state, created by the operator $ S \slashed{\partial} \Psi$.
This is in agreement with \cite{OA}, but there is no
supersymmetric  corresponding fermion-fermion bound state, since the
Gross-Neveu interaction that is responsible for it is absent.

One can also extend the calculation of \cite{AKM,ZZ} to
show that there is no multi-particle production in the heterotic
sigma model. For example, the process $2 \rightarrow 4$ particles can
be shown to vanish.  The reasoning is that the formalism of
\cite{ZZ}, valid for the bosonic case, can be extended to include
superfields.

\section{Critical exponents in the $1/D$ expansion}
\label{EXPONENTS}

\subsection{General discussion}
\label{GENERALITIES}

\begin{figure}[t]
  \centering
  \includegraphics[scale=0.8]{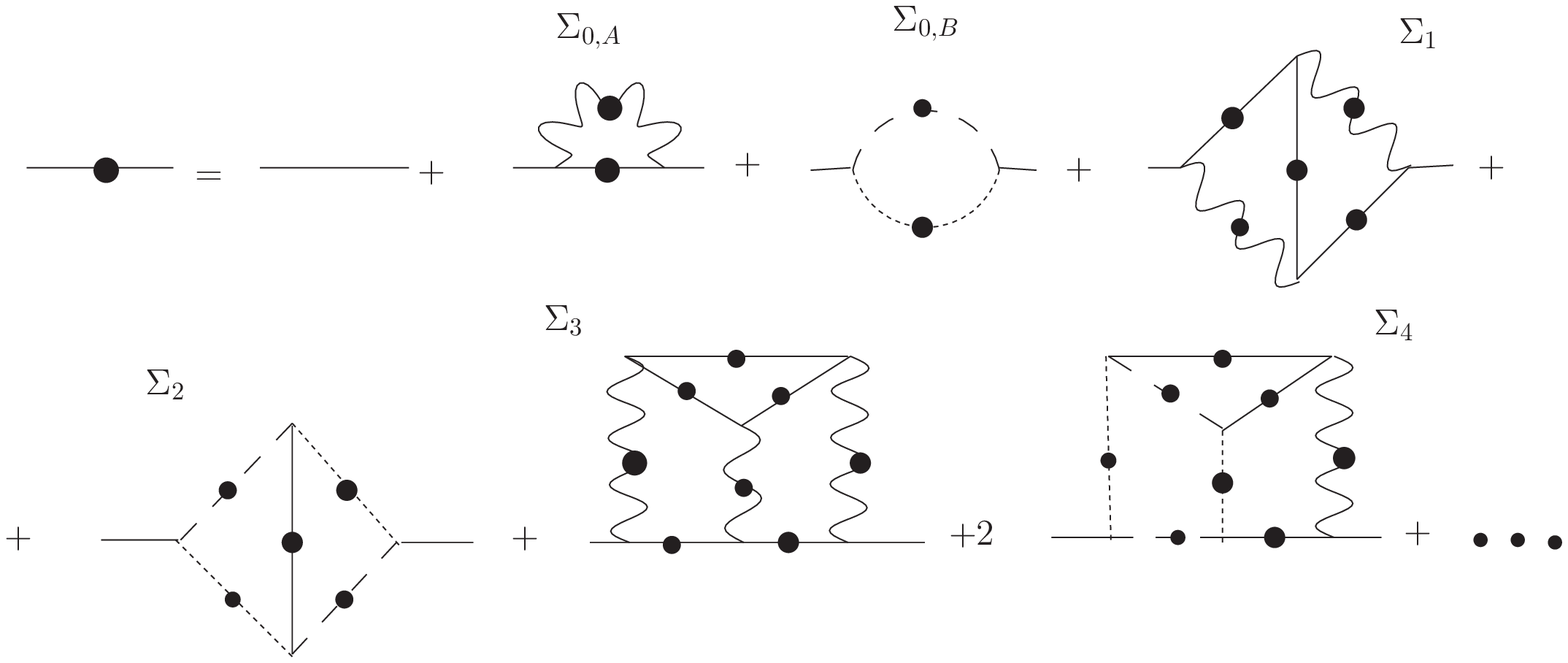}
  \caption{The Dyson equations for the $S$ propagator.}\label{figDysonS}
 \end{figure}

\begin{figure}[t]
  \centering
  \includegraphics[scale=0.8]{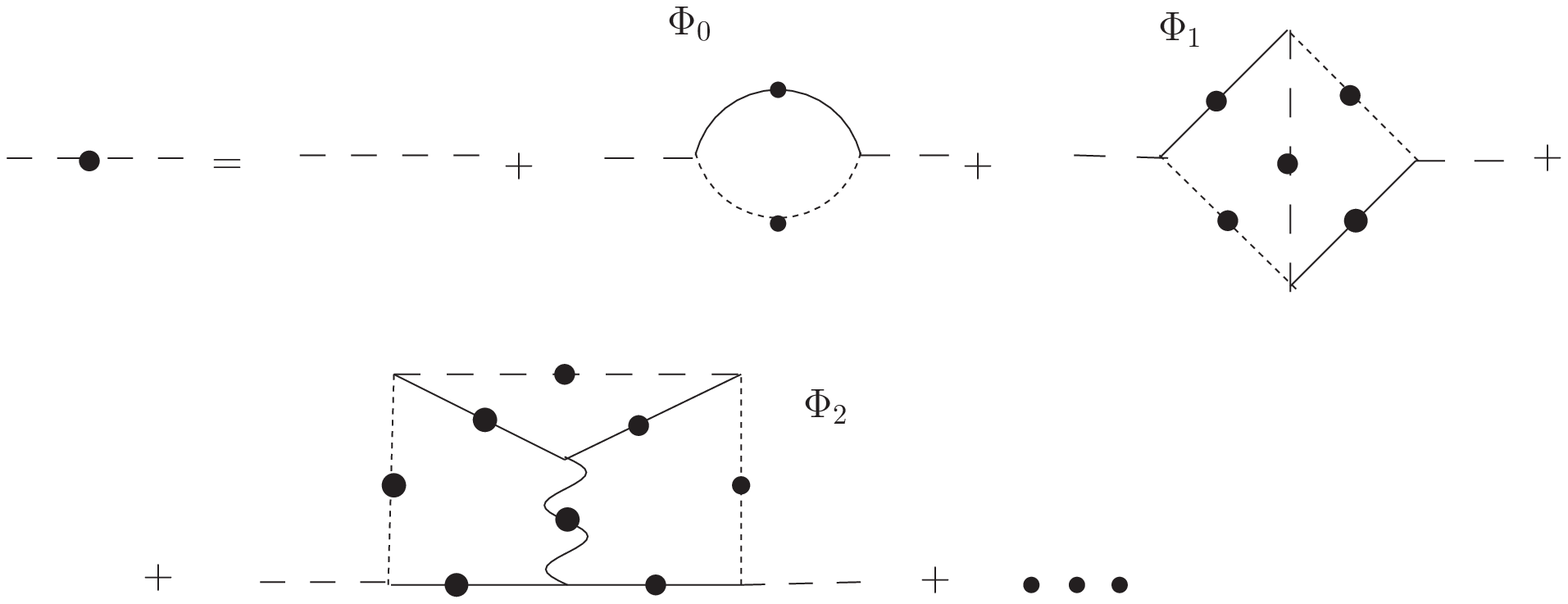}
  \caption{The Dyson equations for the $\Psi$ propagator.}\label{figDysonPsi}
 \end{figure}

\begin{figure}[t]
  \centering
  \includegraphics[scale=0.8]{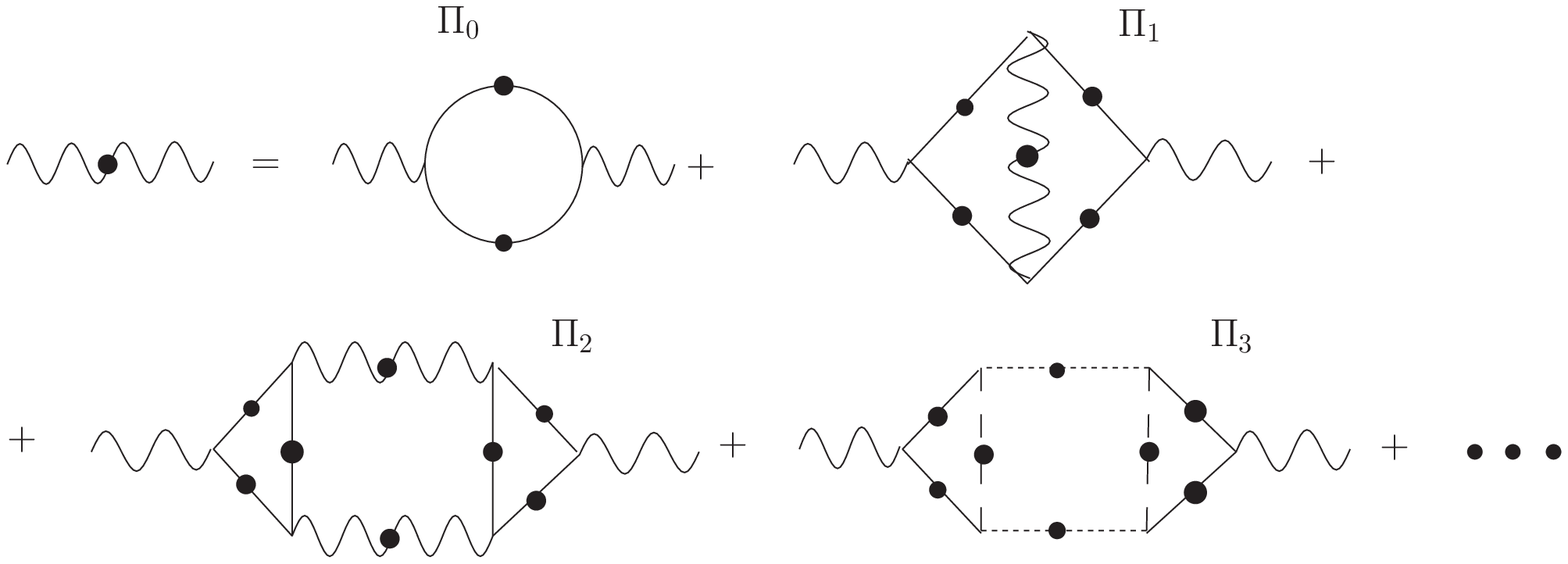}
  \caption{The Dyson equations for the $\sigma$ propator.}\label{figDysonsigma}
 \end{figure}

\begin{figure}[t]
  \centering
  \includegraphics[scale=0.8]{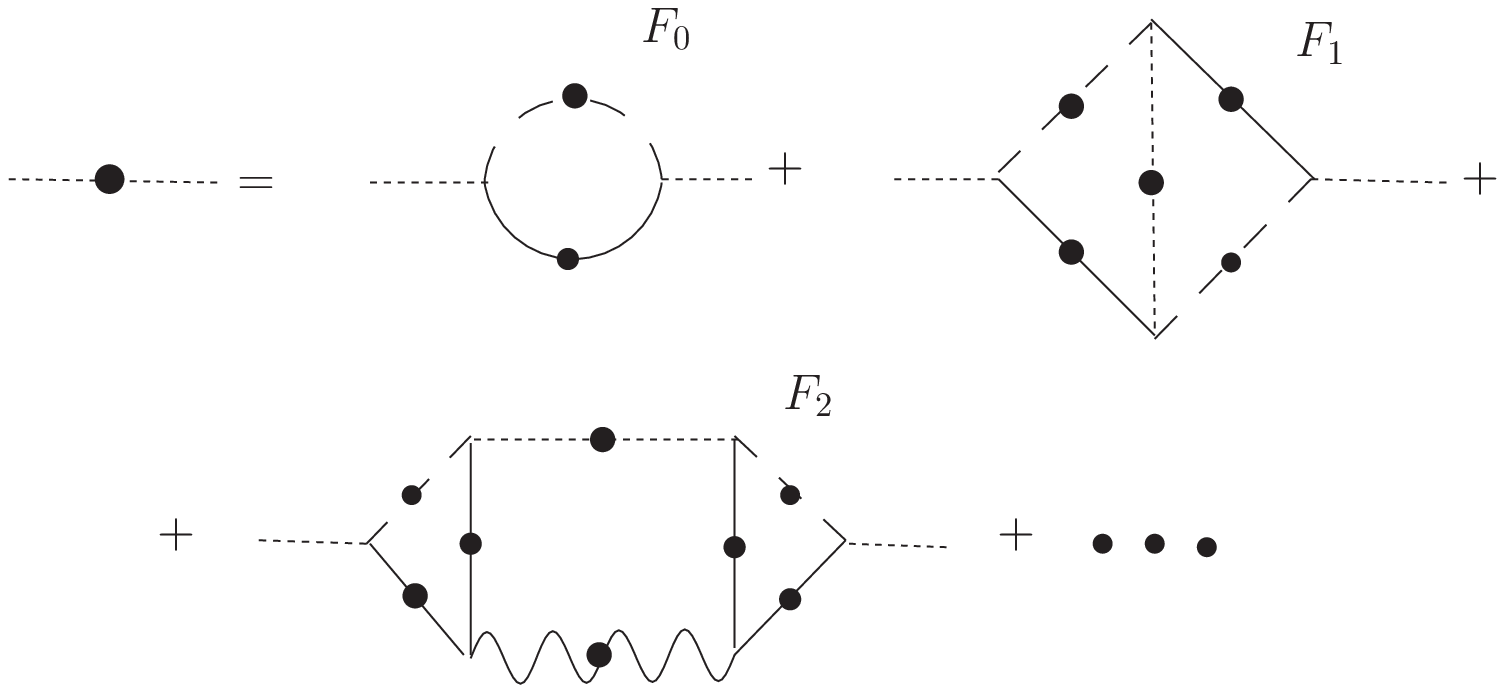}
  \caption{The Dyson equations for the $u$ propagator.}\label{figDysonu}
 \end{figure}

The method used to determine the critical exponents is the one
developed in \cite{Vone,Vtwo} for the bosonic model and
extended to the supersymmetric case in \cite{graceyOne,graceyTwo}. To
this end one writes expressions for the propagators of the fields
near the critical point.
 In keeping with the notation of
\cite{Vone,Vtwo,Ma} we assign dimensions to the fields
\def\dim{\mop{dim}}
 \eqn{AssignDimensions}{
  \dim[S] &= \dim [\Psi]-\frac{1}{2} = \Delta_{S}
    = (d-2+\eta)/2  \cr
  \dim[\sigma] &= \dim[u]+\frac{1}{2} = \Delta_{\sigma}
    = 2 - \eta - \chi \,.
 }
For small but non-zero $x$, the two-point functions may be expanded as follows:
 \eqn{TwoPtFcts}{\seqalign{\span\TL & \span\TR &\qquad \span\TL & \span\TR}{
  G_{SS} (x) &= \frac{\Gamma_{SS}}{x^{2\Delta_{S}}}( 1+\Gamma_{SS}'x^{2\lambda}) &
  G_{\Psi \Psi} (z) &= \frac{1+\gamma_{P}}{2}
    \frac{\Gamma_{\Psi \Psi }\slashed{x}}{x^{2\Delta_{S}+2}}
     (1+\Gamma_{\Psi\Psi}'x^{2\lambda})  \cr
  G_{\sigma \sigma}(x) &= \frac{\Gamma_{\sigma\sigma}}{
    x^{2\Delta_{\sigma}}}(1+\Gamma_{\sigma \sigma}'x^{2\lambda}) &
  G_{uu}(x) &= \frac{1-\gamma_{P}}{2}\frac{\Gamma_{uu}
    \slashed{x}}{x^{2\Delta_{\sigma}}}(1+\Gamma_{uu}'x^{2\lambda}) \,,
 }}
where
 \eqn{GammaPDef}{
  \gamma_{P}=\rho^{0}\rho^{1} =
    \left(\begin{array}{cc} 1 & 0\\ 0 &-1\end{array}\right)
 }
is the chirality matrix in $2$ dimensions.  We have omitted the $O(N)$ indices
because both the propagators and the vertex are proportional to
$\delta_{\mu \nu}$.  So all Green's functions can be expressed as a
scalar function times products of $\delta_{\mu \nu}$, and one does
not have to worry about tensor structures of the form
$(x-y)^{\mu}(x-y)^{\nu}$.  Then one can write the Dyson equations in
 a $1/D$ expansion for the propagators.  Graphical expressions of these equations
 are shown in figures~\ref{figDysonS} through~\ref{figDysonu}.  The Dyson equations
  impose consistency conditions on the critical exponents that determine them completely.
  The graphs that appear in the Dyson equations are the 1PI graphs
with the exception of graphs which contain subgraphs that already
appear in the Dyson equations at a lower effective loop order: in other words, we exclude
diagrams that are already taken into account by expressions for the corrected propagators.
The effective loop order is the number of loops minus the number of loops involving only the
components of the $\Phi$ superfield.

The left hand side of each Dyson equation is a 1PI propagator, which is the inverse of the
connected two-point function.  These inverse propagators are computed by first passing to Fourier space using

 \eqn{inverseFourier}{
   \int d^{d} k  \frac{e^{-ik\cdot x}}{k^{2\Delta}} = \frac{ \pi^{\mu}
  \alpha( \Delta) 2^{2(\mu-\Delta)}}{x^{2(\mu-\Delta)}} \,.
 }
The inverse propagators are found to be\footnote{Note that $G_{\Psi \Psi} \cdot G_{\Psi \Psi}^{-1}$
does not strictly give the unit matrix but instead $\frac{1+\gamma_{p}}{2} =
\left(
\begin{array}{cc} 1&0 \\ 0&0 \end{array}\right)
$ in two dimensions, which is what we want.  Also note that in  the
Dyson equation for $\Psi$ in the right hand side one encounters $u$
propagators that give the right chiral structure, and vice versa for
the Dyson equation of the $u$ field.}
 \eqn{InversePropagators}{
  G_{SS}^{-1}(x) &= \frac{p(\Delta_{S})}{\Gamma_{SS} x^{2(2\mu-\Delta_{S})}}
   (1-q(\Delta_{S},\lambda) \Gamma_{SS}' x^{2\lambda})  \cr
  G_{\Psi \Psi}^{-1}(x) &= \frac{1-\gamma_{P}}{2}
   \frac{p(\Delta_{S})\slashed{x}}{\Gamma_{\Psi \Psi}
    x^{2(2\mu-\Delta_{S})}}(1-s(\Delta_{S},\lambda)\Gamma_{\Psi \Psi}'
    x^{2\lambda})  \cr
  G_{\sigma \sigma}^{-1}(x) &= \frac{p(\Delta_{\sigma})}{\Gamma_{\sigma \sigma}
    x^{2(2\mu-\Delta_{\sigma})}}(1-q(\Delta_{\sigma},\lambda)\Gamma_{\sigma \sigma}'
    x^{2\lambda})  \cr
  G_{uu}^{-1}(x) &= \frac{1+\gamma_{P}}{2}
    \frac{r(\Delta_{\sigma}-1)\slashed{x}}{\Gamma_{uu} x^{2(2\mu-\Delta_{\sigma}+1)}}
     (1-s(\Delta_{\sigma}-1,\lambda)\Gamma_{uu}' x^{2\lambda}) \,,
 }
where
 \eqn{SomeDefs}{
  \mu = d/2 \qquad \Delta_{S} = \mu -1 +\eta/2 \qquad
   \Delta_{\sigma} = 2-\eta - \chi \,,
 }
and for arbitrary $y$,
 \eqn[c]{SomeMoreDefs}{
  \alpha(y) = \frac{\Gamma(\mu-y)}{\Gamma(y)} \qquad
   p(y) = \frac{\alpha(y-\mu)}{\pi^{2\mu} \alpha(y)} \qquad
   r(y) = \frac{y p(y)}{\mu-y}  \cr
  q(y,\lambda) = \frac{ \alpha(y-\lambda)
   \alpha(y+\lambda-\mu)}{\alpha(y)\alpha(y-\mu)} \qquad
   s(y,\lambda) = \frac{y(y-\mu)q(y,\lambda)}{(y-\lambda)
     (y+\lambda-\mu)} \,.
 }
To calculate the beta function, one first evaluates the
critical exponents of the model at the fixed point in $d=2+\epsilon$ dimensions and then uses the relation
\begin{equation}
\lambda = - \frac{1}{2} \beta'(g_{c}) \,,
\end{equation}
valid at the critical point, to extract $\beta(g)$.  This is possible because, in
dimensional regularization with minimal subtraction, the only $\epsilon$ dependence
in $\beta(g)$ is an overall additive term: see \cite{fg} for details.  Expanding in $1/D$ with
$\kappa = gD$ held fixed, one finds
 \begin{equation}\label{betamin}
  \lambda(\epsilon) = \sum_{i=0}^\infty
   \frac{ \lambda_i(\epsilon)}{D^i} \qquad
  \frac{\beta(g)}{g} = \epsilon - \kappa + \sum_{i=1}^\infty
    \frac{ b_i(\kappa)}{D^i}
 \end{equation}
where
 \eqn{GetBs}{
  \lambda_0(\kappa_{c}) = \frac{\kappa_{c}}{2} \qquad
  b_1(\kappa) = - 2\kappa \int_{0}^{\kappa} d \xi
   \frac{\lambda_{1}(\xi)}{\xi^{2}} \qquad
  b_2(\kappa) = -2\kappa \int_{0}^{\kappa} d \xi
   \frac{\lambda_{2}(\xi) - b_{1}(\xi) \lambda'_{1}(\xi)}{\xi^{2}}
    \,.
 }
Note that the critical exponent $\lambda$ is a
measurable quantity and as such it should not depend on the renormalization scheme used.
Passing from the $\lambda_i(\kappa)$ to the $b_i(\kappa)$ does introduce significant scheme dependence.

\subsection{Critical exponents at order $1/D$}
\label{LEADINGORDER}

The Dyson equations can be expressed in terms of parameters
 \eqn{wvDef}{
  w = \frac{\Gamma_{SS}^{2}\Gamma_{\sigma \sigma}}{(2\pi g)^2}
    \qquad
  v = \frac{\Gamma_{SS}\Gamma_{\Psi \Psi} \Gamma_{uu}}{(2\pi g)^2}
    \,,
 }
which can be regarded as dressed vertex factors for the two vertices shown
in figure~\ref{frules}.  The leading non-trivial Dyson equations come from
the graphs labeled $\Sigma_{0,A}$, $\Sigma_{0,B}$, $\Phi_0$, $\Pi_0$, and $F_0$
in figures~\ref{figDysonS} through~\ref{figDysonu}: the tree-level graphs make
no contribution to the leading scaling behavior.  The quantities $\Gamma'_{SS}$,
 $\Gamma'_{\Psi\Psi}$, $\Gamma'_{\sigma\sigma}$, and $\Gamma'_{uu}$ describe how
  far one is removed from the fixed point; so in particular one must be able to
  set them all to zero and get a self-consistent set of equations.  Then the
  dependence of each graph on the position-space separation $x$ is just an overall
  power of $x$.  Matching these overall powers leads simply to the constraint $\chi=0$.
  Matching other factors leads to the equations
 \eqn{LeadingDyson}{\seqalign{\span\TL & \span\TR &\qquad \span\TL & \span\TR}{
  p(\Delta_{S}) + w + v &= 0 &
   r(\Delta_{S}) + v &= 0  \cr
  {1 \over N} r(\Delta_{\sigma}-1) + v &= 0 &
   {2 \over N} p(\Delta_{\sigma}) + w &= 0 \,,
 }}
which determine the quantities $\Delta_S$, $\Delta_\sigma$, $w$, and $v$ as functions of
$\mu$ and $N$.  The system is in fact over-determined if we recall the relations \eno{SomeDefs}
 and the constraint $\chi=0$.  But we will see in section~\ref{ETATWO} that $\chi=0$ is only a
 leading order result; thus to solve \eno{LeadingDyson} we expand
 \eqn{FourExpand}{
  \eta = \sum_{i\geq 0} \frac{\eta_i}{D^i} \qquad
  \chi = \sum_{i\geq 0} \frac{\chi_i}{D^i} \qquad
  w = \sum_{i\geq 0} \frac{w_i}{D^i} \qquad
  v = \sum_{i\geq 0} \frac{v_i}{D^i} \,.
 }
Then one straightforwardly extracts from \eno{LeadingDyson} the coefficients
 \eqn{GotEta}{
  \eta_{0} &= 0 \qquad \chi_0 = 0 \qquad w_0 = 0 \qquad v_0 = 0  \cr
  \eta_1 &= -2
    \frac{\Gamma(2\mu-1)}
      {(\mu-1)^2\Gamma(1-\mu)\Gamma^{2}(\mu-1)\Gamma(\mu+1)}  \cr
  w_1 &= {(2-\mu) \Gamma(\mu-1) \Gamma(\mu+1) \over 2 \pi^{2\mu}}
    \eta_1  \cr
  v_1 &= -{(1-\mu) \Gamma(\mu-1) \Gamma(\mu+1) \over 2 \pi^{2\mu}}
    \eta_1 \,.
 }
Higher order coefficients receive contributions from higher order graphs.  Note that
$\chi_0=0$ could be obtained either from matching overall powers of $x$ or from the equations \eno{LeadingDyson}.

Now consider non-zero coefficients $\Gamma'_{SS}$, $\Gamma'_{\Psi\Psi}$, $\Gamma'_{\sigma\sigma}$,
 and $\Gamma'_{uu}$: this corresponds to moving away from the fixed point.  Linearizing the Dyson equations
 leads to the constraints
 \eqn{LinearConstraints}{
  -p(\Delta_{S})q(\Delta_{S},\lambda) \Gamma'_{SS} +w
   (\Gamma'_{SS}+\Gamma'_{\sigma \sigma}) + v (\Gamma'_{\Psi
   \Psi}+\Gamma'_{uu}) &= 0  \cr
  -r(\Delta_{S})s(\Delta_{S},\lambda)\Gamma'_{\Psi \Psi} +
   v(\Gamma'_{SS}+\Gamma'_{uu}) &= 0  \cr
  -r(\Delta_{\sigma}-1)s(\Delta_{\sigma}-1,\lambda)\Gamma'_{uu} + N
    v(\Gamma'_{SS}+\Gamma'_{\Psi \Psi}) &= 0  \cr
  -p(\Delta_{\sigma}) q(\Delta_{\sigma},\lambda)\Gamma'_{\sigma
   \sigma} + N \Gamma'_{SS} w &= 0 \,.
 }
Graphically, these equations arise from using the leading power-law
 expressions (e.g.~$\Gamma_{SS}/x^{2\Delta_S}$ rather than $G_{SS}(x)$)
 for all propagators except one, chosen arbitrarily; and for that one,
 use the correction term (e.g.~$\Gamma_{SS} \Gamma'_{SS} / x^{2\Delta_S-2\lambda}$).
 The linear equations \eno{LinearConstraints} must admit a non-zero solution for
 $\Gamma_{SS}'$, $\Gamma_{\Psi\Psi}'$, $\Gamma_{\sigma\sigma}'$, $\Gamma_{uu}'$ in
 order for the correction terms to describe a genuine deformation of the critical point.
 So the corresponding determinant must vanish, which leads to
 \eqn{DetCondition}{
  &\left[ (w+v)q(\Delta_{S},\lambda) +
    w \left( 1-\frac{2}{q(\Delta_{\sigma},\lambda)} \right) -
    \frac{v}{s(\Delta_{\sigma}-1,\lambda)} \right]
  (1-s(\Delta_{\sigma},\lambda) s(\Delta_{\sigma}-1,\lambda))  \cr
  &\qquad\quad =
   v\frac{(1-s(\Delta_{\sigma}-1,\lambda))^{2}}
     {s(\Delta_{\sigma}-1,\lambda)} \,.
 }
Note that setting $v=0$ gives the equation valid for the bosonic
model as expected.  To simplify \eno{DetCondition}, one can use
\eno{GotEta} and note that $w_{1}=v_{1} \frac{2-\mu}{\mu-1}$.  So
far we have not used any expansion in $1/D$.  Using the expansions
\eno{FourExpand} and

\eqn{lamdaexpansion}{\lambda = \sum_{i \ge 0}
\frac{\lambda_{i}}{D^{i}}
 }
 we can determine
 \eqn{lambdafirstorder}{
   \lambda_0 = \mu-1 \qquad \lambda_{1} = \frac{1}{2}
   (2\mu-1)(\mu-1)\eta_{1} \,.
 }
 Another way to compute the original determinant, is the following:
 one notes that $r(\Delta_S)s(\Delta_S,\lambda) \sim 1/D^0$, while all other
terms scale at least as $1/D^{1}$.  This means that, in expanding
the $4\times4$ determinant of \eno{LinearConstraints}, the first
order contribution comes only from the determinant of the $3 \times
3$ matrix
 \eqn{AIdef}{
  AI \equiv \left( \begin{array}{ccc} -p(\Delta_{S})q(\Delta_{S},\lambda) +w & w&v \\
 w & -p(\Delta_{\sigma})q(\Delta_{\sigma},\lambda)/N & 0 \\
  v & 0 & -r(\Delta_{\sigma}-1)s(\Delta_{\sigma}-1,\lambda)/N
\end{array} \right) \,.
 }
 where in each element of the matrix we only keep the first term in
 the $ 1/D$ expansion.   This determinant provides the first  $1/D $ term  of \eno{DetCondition} and thus reproduces  \eno{lambdafirstorder}.

 It is interesting to compare with the bosonic case.  It is
easily seen that $\lambda_1^{\rm het} = {1 \over 2} \lambda_1^{\rm
bos}$, even though $\eta^{\rm het}_1$, $w_1$, and $v_1$ are not so
simply related to $\eta^{\rm bos}_1$
 and the corresponding vertex factor for the bosonic case.  The relation
 $\lambda_1^{\rm het} = {1 \over 2} \lambda_1^{\rm bos}$ is expected: we know
 that $\lambda_{1}^{\rm sup} =0$ in the type~II case, and having
half the fermions in the heterotic case will cancel only half the bosonic contribution.

\subsection{A check of the calculation}
\label{CHECK}

 Following \cite{Ma}, we see that $\eta$ is the anomalous dimension
 of the $S$ propagator.  So far we have computed it in the $1/D$ expansion
 using techniques in position space.  One can also straightforwardly
 compute $\eta_{1}$ in momentum space using the expressions for the
 propagators that we have previously found \eno{uprop}.  Firstly
 we note that for small $k$,
 \eqn{smallkapppa}{
 \tilde{G}_{SS}(k) \sim k^{-2+\eta} \sim
 k^{-2+\eta_{o}}\left(1+\frac{\eta_{1}}{N} \log k +
 \mathcal{O}(1/N^{2}) \right) \,,
 }
with $\tilde{G}_{SS}$ the Fourier transform of the $S$
 propagator.  But $\tilde{G}_{SS}(k) $ can also be determined from the one-particle irreducible diagrams $\Sigma(k^{2})$:
 \eqn{sigmamomentum}{
 \tilde{G}_{SS}^{-1}(k) = k^{2} +\Sigma(k^{2}) - \Sigma(0) \sim
 k^{2}(1-\frac{\eta_{1}}{N} \log k ) \,,
 }
since $\eta_{0}=0$.  Having calculated the propagators of the
lagrange multiplier fields
 it is straightforward to compute the $\Sigma(k^{2})$ from the
 one-loop diagrams.  We find
 \eqn{sigmaintegral}{
 \Sigma (k^{2}) &= \int
 \frac{d^{d}p}{(2\pi)^{d}} \frac{i}{(p-k)^{2} }D_{u}(p^{2}) - \int
 \frac{d^{d} p}{(2\pi)^{d}} \tr \frac{i}{(\slashed{k}-\slashed{p})}
 S_{u}(\slashed{k})  = \frac{2}{N}\int \frac{
 d^{d}p}{(2\pi)^{d}} \frac{ \tr(\slashed{k}
 \slashed{p})V(p^{2})}{(p+k)^{2}}  \cr
   &=
 \frac{2^{d-2}/N\sqrt\pi}{_{2}F_{1}(2-d/2,1/2,3/2,1)\Gamma(2-d/2)\Gamma(\frac{d-1}{2})}
  \int _{0}^{ M} d p \int_{0}^{\pi} d \theta \frac{ p^{2} k \cos
 \theta \sin^{d-2} \theta }{ p^{2} +k^{2} +2pk \cos \theta}
  }
 where
we have put $m^{2}=0$ as usual \cite{graceyTwo}, and $M$ is the cutoff.
One notes that, for small $k$ the $k^{2}\log k$ behavior comes from the
small $p$ region \cite{Ma}.  The integral is trivial to do, and for the
$k^{2} \log k$ part it gives
 \eqn{sigmaresult}{
 \Sigma(k^{2})=-\frac{2^{d-1}}{N d} \frac{1}{_2F_{1}(2-d/2,1/2,3/2,1)
 \Gamma(2-\frac{d}{2})\Gamma(\frac{d}{2})} k^{2} \log k \,.
 }
 It is easy to see using $d=2 \mu$ and properties of the Gamma function that
this coincides with the expression for $\eta_1$ in \eno{GotEta}.  An
easier way to do the checking is the following. One can write
similar expressions for $\Sigma(k^{2})$ for the bosonic and
supersymmetric models \cite{Vone,graceyTwo}. Then it is easy to
observe that
  \eqn{sigmarelation}{
  \Sigma_{\rm bos}(k^{2})
  -\Sigma_{\rm bos}(0) = -\Sigma_{\rm sup}(k^{2}) +2\Sigma_{\rm
   het}(k^{2}) \,.
   }
 This
easily gives\footnote{ We have used that $\Sigma_{\rm sup}(0) =
0$ and $\Sigma_{\rm het}(0) =0$. }
  \eqn{etarelations}{
  2\eta_{\rm het} = \eta_{\rm bos}+\eta_{\rm sup} \,.
  }
 With \cite{Vone,graceyTwo}
  \eqn{etaCompare}{
 \eta_{\rm bos} =\frac{(2-\mu)}{\mu } \eta_{\rm sup}  \qquad
 \eta_{\rm sup} =
 \frac{4}{N}\frac{\Gamma(2\mu-2)}{\Gamma^{2}(\mu-1)\Gamma(2-\mu)\Gamma(\mu)}
 }
we
find $ \eta_{\rm het} = \frac{1}{\mu}\eta_{\rm sup}$ which agrees
with \eno{GotEta}.

\section{Results at order $1/D^2$}
\label{NEXTORDER}

Each graph in figures~\ref{figDysonS}-\ref{figDysonu} carries an overall
 factor $1/D^M$ where $M$ is the number loops minus the number of loops
 containing only $S$ and $\Psi$.  To see this, first note that each propagator $G_{XX}$
 carries a factor $\Gamma_{XX}$ (where $X = S$, $\Psi$, $\sigma$, or $u$).
  Next note that the amplitude for each graph must contain an overall factor
  which is a product of the factors $w = \Gamma_{SS}^2 \Gamma_{\sigma\sigma} / (2\pi g)^2$
  and $v = \Gamma_{SS} \Gamma_{\Psi\Psi} \Gamma_{uu}/(2\pi g)^2$, one for each vertex
  in the graph.  The overall factor $1/D^M$ arises because $w$ and $v$ scale as $1/D$
  and because each loop containing only $S$ and $\Psi$ carries a factor of $N$.
  The graphs in figures~\ref{figDysonS} and~\ref{figDysonPsi} are those with $M \leq 2$,
  and the ones in figures~\ref{figDysonsigma} and~\ref{figDysonu} are those with $M \leq 1$.
   Together, these are all the graphs that can contribute to $\eta$, $w$, $v$, and $\lambda$
    through order $1/D^2$, and they also determine $\chi$ through order $1/D$.  Because we
    quote final results in terms of $1/D$, we must keep in mind the relation between expansions
     in $1/N$ and $1/D$:
 \eqn{wExpand}{
  w = \sum_{i \geq 0} {\tilde{w}_i \over N^i} =
    \sum_{i \geq 0} {w_i \over D^i} \qquad
  w_1 = \tilde{w}_1 \quad w_2 = \tilde{w}_2 - 2 \tilde{w}_1 \,,
 }
with similar relations for other quantities.

\subsection{Calculation of $\eta_2$}
\label{ETATWO}

A technical complication arises in the $1/D^2$ corrections to the Dyson equations that
 was explained and resolved in \cite{Vone,Vtwo}.  The problem is that the higher-loop
 graphs diverge when $\chi=0$.  In fact, $\chi=0$ only up to $1/D$ corrections.  But it convenient to
 regularize the ``divergence'' and extract finite expressions for the two-loop Dyson equations through
 the following steps:
 \begin{enumerate}
  \item Shift $\chi \to \chi+\Delta$.
  \item Expand the amplitudes for individual graphs in powers of $\Delta$.
  \item Cancel $1/\Delta$ terms against certain counter-terms in the lagrangian.
  \item Fix $\chi_1$ by setting to zero certain terms in the Dyson equation which depend logarithmically on
  the position-space separation $x$ and which, if non-zero, would spoil self-consistency.
 \end{enumerate}
We will now go through these steps in detail for the $\Psi$ propagator.  The reader who wishes to bypass
the technical details can skip to \eno{DysonPhi} and \eno{DysonSSigmaU}, which are the two-loop Dyson equations
with all divergences removed.  But the results \eno{ChiOneDeterm} and \eno{mazaSeveral} for $\chi_1$ provide
important consistency checks.

When $\chi \neq 0$, $x$ dependence cannot be canceled out of the Dyson equations in a simple way: setting
   $\Gamma'_{SS} = \Gamma'_{\Psi\Psi} = \Gamma'_{\sigma\sigma} = \Gamma'_{uu} = 0$,
   one obtains for the $\Psi$ propagator's Dyson equation
 \eqn{dyspsi}{
  r(\Delta_{S}) + v (x^{2})^{\chi} +
   v^{2}(x^{2})^{2\chi} \Phi_1 + N
   w v^{2}(x^{2})^{3\chi}\Phi_2 = 0 \,.
 }
Here $\Phi_1$ and $\Phi_2$ are functions of $\Delta_S$, $\Delta_\sigma$, and $\mu$
which diverge when $2\Delta_{S}+\Delta_{\sigma} -2\mu = -\chi=0$.  Although these
are in some sense an artifact of a limit ($\chi \to 0$) which one cannot take
independently of the large $N$ limit, it is convenient nevertheless to regulate them, as explained above, by shifting
 \eqn{ShiftChi}{
  \chi \to \chi + \Delta \,.
 }
The amplitudes $\Phi_{1,2}$ may then be expanded as
 \eqn{ExpandPhis}{
  \Phi_i = {X_i \over \Delta} + \Phi_i' + O(\Delta) \,,
 }
where both $X_i$ and $\Phi_i'$ are functions of $\Delta_S$, $\Delta_\sigma$, and
$\mu$, subject to $2\Delta_S+\Delta_\sigma-2\mu=0$.
 In appendix~\ref{ETADETAILS} we exhibit $\Phi_{1,2}$ in the form \eno{ExpandPhis}, as well
  as a number of related quantities that enter into other Dyson equations.  To cancel the
  divergent $1/\Delta$ terms in $\Phi_{1,2}$, one may rescale the lagrange multiplier fields
  in the original action \eno{hetaction}.  This rescaling amounts to adding counter-terms to the
  action, and it can be expressed, to the relevant order, as
 \eqn{RescaleVW}{
  v \to \left( 1 + {m_1 \over N} \right) v \qquad
  w \to \left( 1 + {m_1 \over N} \right) w \,.
 }
(The factor on $v$ and $w$ is the same because of supersymmetry.)
Subjecting \eno{dyspsi} to the shift \eno{ShiftChi} and the
rescaling \eno{RescaleVW}, it becomes, keeping terms up to $1/N^{2}$,
  \eqn{dyspsibec}{
   & r(\Delta_{S}) + (x^{2})^{\chi}\Big( v+ v^{2} \Phi'_{1} + N v^{2} w
   \Phi'_{2} \Big) + \cr
     &  (x^{2})^{\chi}\Big( v_{1}\frac{m_{1}}{N} + v_{1}^{2}
   (x^{2} )^{\chi} \frac{X_{1}}{\Delta} + N v_{1}^{2} w_{1}
   (x^{2})^{2\chi}
   \frac{X_{2}}{\Delta}\Big) =0
 }
 The last line contains all the divergent pieces. Setting $ \chi = 0
 $ \cite{Vone} and taking the limit $\Delta \to 0$ determines $m_{1}$ as
 \eqn{MassOne}{
-v_{1} \frac{ m_{1}}{N} = v_{1}^{2}
    \frac{X_{1}}{\Delta} + N v_{1}^{2} w_{1}
   \frac{X_{2}}{\Delta} \,.
   }
Plugging \eno{MassOne} back into \eno{dyspsibec}, and now
considering a finite $\chi$ we get
\eqn{dyspsionemore}{
 r(\Delta_{S}) + (x^{2})^{\chi}\left( v+ v^{2} \Phi'_{1} + N v^{2} w
   \Phi'_{2} \right) + v_{1}^{2}X_{1}\left(
   \frac{(x^{2})^{2\chi}-(x^{2})^{\chi}}{\chi}\right)&  \cr
   {} + N
   v_{1}^{2}w_{1}X_{2}\left( \frac{(x^{2})^{3\chi}-
   (x^{2})^{\chi}}{\chi}\right) =0& \,.
   }
When one expands $\chi = \chi_1/N + O(N^{-2})$, there are terms that
behave as $\log x^{2}$. One gets rid of these if $\chi_{1}$ obeys
  \eqn{ChiOneDeterm}{
  \chi_{1} = -v_{1}X_{1}- 2 v_{1}w_{1}X_{2} \,.
  }
We could have derived \eno{MassOne},\eno{ChiOneDeterm} purely within the $N \to \infty $ limit.
 In this setup there is no need  for $\Delta$ and $\chi$ is taken to be finite. Note that it
 behaves as $\chi \sim 1/N$ since $\chi_{0}=0$ \eno{GotEta}. The Dyson equation after the
 rescaling \eno{RescaleVW} is \eno{dyspsibec} with $\Delta$ replaced with $\chi$. In taking
 the $ N \to \infty$ limit there are terms that diverge  linearly with $ N$ and terms that behave
 as $ \log x^2$. Respectively these are
  \eqn{DysonDiverg}{
  (x^2)^{\chi}\big(v_{1}\frac{m_{1}}{N} + v_{1}^{2}\frac{X_{1}}{\chi}+Nv_{1}^{2} w_{1}\frac{X_{2}}{\chi}\big) \cr
  \log x^{2} \big(\chi  v_{1}(1+\frac{m_{1}}{N}) + 2 v_{1}^{2} X_{1} + 3 N v_{1}^{2} w_{1} X_{2} \big) \,.
  }
Setting these to zero fixes $m_{1},\chi_{1}$ as in \eno{MassOne},\eno{ChiOneDeterm}, with $\Delta$
 replaced by $\chi$.
  We choose to keep  the $\Delta$ shift, as is common in the literature \cite{Vone,Vtwo,graceyThree, graceyFour}.
   If one wishes to translate our results, in the  $N \to \infty $ formalism, only a  simple substitution of
   $ \Delta \to \chi_{1}/N $
   is needed in the values of the diagrams given in Appendix \ref{ETADETAILS}.  What is
 left is the finite correction to the leading Dyson equation for the $\Psi$ propagator:
 \eqn{DysonPhi}{
  r(\Delta_S) + v + v^2 \Phi'_1 + N w v^2 \Phi'_2 = 0 \,.
 }
Following the same procedure for the $S$, $\sigma$, and $u$ Dyson equations, one gets the finite equations
 \eqn{DysonSSigmaU}{
  p(\Delta_S) + w + v + w^2 \Sigma'_1 - v^2 \Sigma'_2 +
    N w^3 \Sigma'_3 -2N v^2 w \Sigma'_4 &= 0  \cr
  p(\Delta_\sigma) + {N \over 2} w + {N \over 2} w^2 \Pi'_1 +
  \frac{N^{2}}{2} w^{3} \Pi_{2}' -
   \frac{N^{2}}{2} wv^{2} \Pi_{3}' &= 0  \cr
  r(\Delta_{\sigma}-1) + Nv + N v^{2} F_{1}' +
   N^{2} wv^{2} F_{2}' &= 0 \,,
 }
where $\Sigma'_i$, $\Pi'_i$, and $F'_i$ are the finite parts of $\Sigma_i$, $\Pi_i$, and $F_i$,
listed in Appendix~\ref{ETADETAILS}.  The minus signs in \eno{DysonSSigmaU} come from fermion
 loops.  From each Dyson equation one also
  gets a new determination of $m_1$ and $\chi_1$:
 \eqn{mazaSeveral}{\seqalign{\span\TL & \span\TR &\qquad \span\TL & \span\TR}{
  m_1 &= {1 \over \Delta}
   {w_{1}^{2} S_{1} -v_{1}^{2} S_{2} + w_{1}^{3} S_{3} -
     2 w_{1}v_{1}^{2} S_{4} \over w_1 + v_1} &
  \chi_1 &= {-w_1^2 S_1 + v_1^2 S_2 - 2 w_1^3 S_3 +
    4 w_1 v_1^2 S_4 \over w_1+v_1}  \cr
  m_1 &= -{1 \over \Delta} (w_1 P_1 + w_1^2 P_2 - v_1^2 P_3) &
    \chi_{1} &= -w_1 P_1 - 2w_1^2 P_2 + 2 v_1^2 P_3  \cr
  m_{1} &= -\frac{1}{\Delta} ( v_{1} Y_{1} + w_{1}v_{1}Y_{2}) &
    \chi_{1} &= - v_{1} Y_{1} -2 w_{1}v_{1} Y_{2} \,,
 }}
where $P_i$, $S_i$, $Y_i$ are the residues of $\Pi_i$, $\Sigma_i$, and $F_i$, respectively.

Fortunately, the four seemingly independent determinations of $m_1$ and $\chi_1$ all agree,
 as one can check by explicitly evaluating \eno{MassOne}, \eno{ChiOneDeterm}, and \eno{mazaSeveral}
  using expressions from Appendix~\ref{ETADETAILS} with $\Delta_S=\mu-1$ and $\Delta_\sigma=2$.
  This provides a check that the renormalization procedure we have chosen to cancel the divergences
  of higher-loop graphs is consistent.  Other schemes change the values for individual amplitudes,
   but the critical exponents remain the same \cite{graceyOne}.

Interestingly, there is yet another consistency check on $\chi_1$.  One can show from
\eno{ChiOneDeterm} or \eno{mazaSeveral} that
 \eqn{FinalChiOne}{
  \chi_{1} = \mu(2\mu-3) \eta_{1} \,.
 }
This is seen to comply with a scaling law formulated for the bosonic model in \cite{Ma}:
 \eqn{scalingma}{
  2\lambda = 2\mu -\Delta_\sigma \,.
 }
That this relation is also valid in our case can be seen by applying the Callan-Symanzik
equation near the critical point for $\langle
\sigma(p)\sigma(-p)\rangle$ or $\langle\bar{u}(p) u(-p)\rangle $
propagator.

Now we can solve \eno{DysonPhi}-\eno{DysonSSigmaU} by
eliminating $w$ and $v$:
  \eqn{dysetatwo}{
  r(\Delta_{S}) =
  \frac{1}{N}r(\Delta_{\sigma}-1) + \frac{v_{1}^{2}}{N^{2}} ( F_{1}'-
  \Phi_{1}') + \frac{w_{1}v_{1}^{2}}{N^{2}}( F_{2}' - \Phi_{2}') \,.
  }
 Expanding $\Delta_{S},\Delta_{\sigma}$ in \eno{dysetatwo}, we can determine
$\tilde{\eta_{2}}$
  \eqn{Etatwo}{
 \frac{\tilde{\eta}_{2}}{\eta_{1}^{2}} &=\frac{1}{2\mu}
 +(\mu-1)(2\mu-1)\left(-1+\pi \cot \mu \pi +H(2\mu-2)\right) \cr
 + \frac{\mu}{\mu-2} + \frac{1}{2(\mu-1)}& -1
  -\mu(\mu-2)\left(B(2\mu-3)-B(\mu-1)
  -\frac{1}{\mu-1}+\frac{1}{2\mu-3} -2\right)
  }
 where $H(x)= \psi(x+1) -\psi(1) $  and the $B(x)$ function is defined in the
 appendix. The first line  just comes from the Hatree-Fock diagrams, i.e. by iterating the 1/N
Dyson equation to the next order, the second is the contribution of
$\Phi_{1},F_{1}$ and the third comes from $ \Phi_{2},F_{2}$. We also
can determine the values of $ w_{2},v_{2}$ as
 \eqn{wTwo}{
  \frac{\tilde{w}_{2}}{\eta_{1}w_{1}} &= \frac{(2
  \mu-1)(\mu-1)}{\mu-2}\left(3-\mu + (\mu-2)  \pi \cot
\mu \pi + (\mu-2)H(2\mu-3) \right)\cr &\qquad{}
    -\mu \left( (7\mu-9) B(\mu-1) + (13-10\mu) B(2) +
    (3\mu-4)B(2\mu-3) \right) \cr
   &\qquad\quad{} - \mu + 2 \mu(\mu-1)-\frac{\mu(\mu-1)}{2\mu-3}
 }
 \eqn{vTwo}{
 \frac{\tilde{v}_{2}}{\eta_{1}v_{1}} =
\frac{\tilde{\eta_{2}}}{\eta_{1}^{2}} -\frac{1}{2\mu}
  -\frac{\mu}{2} +\mu(\mu-2) +2\mu(\mu-2)\left(B(2)-B(\mu-1)\right)  \,.
 }
In \eno{wTwo}, the first three terms come from iteration of the first order equations, while in \eno{vTwo},
the first two terms come from such iteration.

\subsection{Calculation of $\lambda_2$}
\label{LAMBDATWO}

As in section~\ref{LEADINGORDER}, the calculation of $\lambda_2$ through order $1/D^2$ requires
evaluating each graph with one propagator altered from its leading power behavior
(e.g.~$\Gamma_{SS}/x^{2\Delta_S}$ for an $S$ propagator) to its sub-leading power behavior
(e.g.~$\Gamma_{SS} \Gamma'_{SS} / x^{2\Delta_S-2\lambda}$).  The four Dyson equations lead to
four linear equations in the quantities $\Gamma'_{SS}$, $\Gamma'_{\Psi\Psi}$, $\Gamma'_{\sigma\sigma}$,
and $\Gamma'_{uu}$:
 \eqn{LS}{
  (-p(\Delta_{S})q(\Delta_{S},\lambda)  + w+\Sigma_{S}) \Gamma_{SS}' +
   (w+\Sigma_{\sigma}) \Gamma_{\sigma \sigma}' +
   (v+\Sigma_{u})\Gamma_{uu}' + (v+\Sigma_{\Psi}) \Gamma_{\Psi \Psi}'
   &= 0  \cr
  (-r(\Delta_{S})s(\Delta_{S},\lambda)+ \Phi_{\Psi})
   \Gamma_{\Psi\Psi}' + (v+ \Phi_{S}) \Gamma_{SS}' +( v+ \Phi_{u})
   \Gamma_{uu}' &= 0  \cr
  \left( -\frac{p(\Delta_{\sigma})q(\Delta_{\sigma},\lambda)}{N}+
    \Pi_{\sigma} \right) \Gamma_{\sigma\sigma}' +
    (w+\Pi_{S})\Gamma_{SS}' + \Pi_{\Psi} \Gamma_{\Psi\Psi}' +
    \Pi_{u} \Gamma_{uu}' &= 0  \cr
  \left( -\frac{1}{N}r(\Delta_{\sigma}-1)
    s(\Delta_{\sigma}-1,\lambda) + F_{u} \right) \Gamma_{uu}' +
    (v+F_{S})\Gamma_{SS}'+(v+F_{\Psi})\Gamma_{\Psi\Psi}' &= 0 \,,
 }
where for example we denote by $\Sigma_\Psi$ all the diagrams
that appear in the $S$ propagator where the $\Psi$ propagator is
corrected.  As in section~\ref{ETATWO}, the amplitudes diverge when
$2\Delta_S + \Delta_\sigma - 2\mu \to 0$, and the same procedure described there to regulate and
subtract the divergences and to remove terms proportional to $\log x^2$ carries over to the present
case.  The finite parts of all the quantities in \eno{LS} are given in Appendix~\ref{LAMBDADETAILS},
as well as some further remarks on their evaluation.

The system \eno{LS} must have a nonzero solution for the
$\Gamma$'s, so the determinant must be zero.  This determines
$\lambda_2$.  A way to calculate the determinant to sufficient
accuracy is to note that $r(\Delta_{S}) s(\Delta_{S},\lambda) \sim
1/N^{0}$, and then expand the determinant into three $3 \times 3$
 determinants, i.e.~expanding  in the line of $\Psi$ field Dyson equation.  All terms have to be expanded up
 to $1/N^2$ accuracy. One also
 notes that $\lambda_{2}$ only appears in the expansion of $
 p(\Delta_{S})q(\Delta_{S},\lambda)$ at this order. So $\lambda_{2}$
 is going to be a linear combination of the various sums of
 diagrams given in the appendix, factors of $w_{2}$ and $v_{2}$,  and terms
 that come from iterating the $1/N$ equations.  The final result is quite
 involved and we prefer to give it implicitly as
 \eqn{lambdatwo}{
  {\tilde\lambda_2 \over \eta_1^2} &=
  -\frac{1}{2(\mu-1)^2}-\frac{25}{2(\mu-1)}+25-\frac{5(\mu+1)}{2(\mu-2)}-\frac{5}{4(2\mu-3)}-\frac{\mu-2}{2}
  +50(\mu-1)
   \cr
  &\qquad{}+(\mu-1)\left(-\frac{19}{8}(\mu-2) +\frac{45}{2}(\mu-2)^2-\frac{3}{2}(2\mu-3)\right) -
   \frac{\mu^{2}(2\mu-3)^{2}}{8(\mu-1)}
      \cr
   &\qquad{}+(\mu-1)^2 \left( \frac{21}{4}-\frac{9}{4}(\mu-2)-75(2\mu-3)-\frac{25}{(\mu-2)} - \frac{10}{(2\mu-3)}
    \right)
   \cr
  &\qquad{} -2\mu(\mu-1)\tilde{v}_{2}-
   \left( \frac{\mu}{2}+2\mu(2\mu-3) \right) \tilde{w}_{2}
    -2\mu(\mu-1)(\mu-2)\frac{F_{u}+F_{S}}{\eta_{1}v_{1}}  \cr
  &\qquad{} -\mu(\mu-1)\frac{(2\mu-3)^{2}}{\mu-2}
   \frac{\Pi_{\sigma}}{\eta_{1}v_{1}}
    -2\frac{\mu(\mu-1)}{(\mu-2)^{2}}\frac{\Sigma_{S}}{\eta_{1}v_{1}}
    -\mu \left( \frac{\mu-1}{\mu-2}-2\mu \right)
     \frac{\Sigma_{\sigma}}{\eta_{1}v_{1}}  \cr
 &\qquad{}
    -3\mu(2\mu-3)(\mu-1)\frac{F_{\sigma}}{\eta_{1}v_{1}} -\frac{3}{2}\left(\pi \cot \mu\pi
    +H(2\mu-4)\right)\,.
 }
The right hand side is a function of $\mu$ which can be obtained
explicitly by substituting the expressions
\eno{GotEta}, \eno{wTwo}, \eno{vTwo}, \eno{SigmaSsum},
\eno{Sigmasigmasum}, \eno{Sigmausum}, \eno{Pisigmasum},
 \eno{Piusum}, and \eno{Fusum} into \eno{lambdatwo}.

\subsection{Calculation of the beta function}
\label{BETA}

As explained in section~\ref{GENERALITIES}, we can calculate the beta function once we know $\lambda$.
 Noting that \eno{lambdatwo}
 gives the $1/N^{2}$ expansion term and subtracting $2\lambda_{1}$ we find
 \eqn{lones}{
 \lambda_{0} = \epsilon/2 ,\quad  \lambda_{1}(\epsilon)
 =\frac{\epsilon^{2}}{4}+\frac{\epsilon^{3}}{8}
- \frac{\epsilon^{4}}{16} +\mathcal{O}(\epsilon^{5})
 }
\eqn{ltwos}{
  \lambda_{2} =
 \left(\frac{5}{16}+\frac{9}{4}\zeta(3)\right)\epsilon^{4}
 +\mathcal{O}(\epsilon^{5}) \,.
 }
 Using \eno{GetBs}, we compute the beta function for the heterotic
 string in a constant curvature background:
 \eqn{betafunction}{
 \beta(g) = -Dg^{2}-\frac{1}{2}D g^{3}
 -\frac{g^{4}D}{4}\left(1+\frac{D}{2}\right)-\frac{g^{5}D^{2}}{4}\left(\frac{3}{2}-\frac{D}{3}\right)
 -\frac{3}{2}\zeta
 (3) g^{5} D^{2} + \mathcal{O}(\frac{1}{D^{3}}) \,.
 }
 It is obvious that there is agreement with the first two
loops of the expression \eno{LeadingCorrection}, where we use
 \eqn{BetaRelation}{
 \beta(g) = M \frac{\partial g}{\partial M} = - \frac{g}{N-1}
  g^{ij}\beta_{ij}
 }
and \eno{AdScurvature}. We do not know of any calculation of the
beta function of the heterotic string in the minimal subtraction
scheme that goes beyond two loops.  In \cite{FMRone,FMRtwo} the
beta function was computed using the background field method, and
found to be in three loops
 \eqn{FMRresult}{
  \beta_{ij}^{(3)} = \frac{\alpha'^3}{8}\left( \frac{3}{2} R_{ikjl}
  R^{kmnp} R^{l}{}_{mnp} -\frac{1}{2}R_{lm}R_i{}^{lnp}R_j{}^{m}{}_{np} -
  \frac{1}{2}R_{jl}R^{lmnp}R_{imnp}\right) \,.
  }
The
appearance of the Ricci tensor means that it is not minimal
subtraction.  Divergences involving the Ricci tensor can only appear
through closed loops where at least one propagator starts and ends at the
same vertex.  Within the minimal subtraction scheme, at more than
one loop these terms  combined with their counterterms never produce
a simple pole \cite{grisaruOne,grisaruTwo}.  A small check of our
result comes from the famous $\zeta(3)$ term, $ \frac{\zeta(3)
   \alpha'^4}{2} R_{mhki}R_{jrt}{}^m (R^k{}_{qs}{}^r R^{tqsh} +
     R^k{}_{qs}{}^t R^{hrsq})$.  This term is identical in the bosonic \cite{Wegner,Jack:1988sw,Jack:1989vp}, supersymmetric \cite{grisaruOne,grisaruTwo}, and heterotic \cite{gross-sloan} cases.  In an expansion of the Virasoro amplitude, it is associated with the constant term in an expansion in the Mandelstam variables $s$, $t$, and $u$.  At loop order $n+1$ in \NLsM\ calculations, it seems likely that the coefficient of $\zeta(n)$ is the same for the bosonic, supersymmetric, and heterotic cases (see \cite{graceyOne} for a comparison of the bosonic and supersymmetric cases).

In \cite{MT,gross-sloan,CN}, the absence of a three-loop term of the form $\alpha'^3 R^{3}$ was noted.  The three-point scattering
amplitudes suggest that there are also no $RF^{2}$ or $F^{3}$ in the
effective action.  One knows that identifying the gauge connection
with the spin connection in the heterotic string effective action
will give the superstring effective action, where there is no
$\alpha'^3$ term.  So if there were any $R^{3}$ terms in the
heterotic case it would not be possible to cancel them.  All this seems in conflict with the \eno{betafunction}, where the term proportional to $g^4$ would seem to correspond to an $R^3$ term in the effective action.  But it
should be noted that the relation between the effective action and
the beta function is \cite{Tseytlin:1986ti}
 \eqn{Seffts}{
 2\kappa_{D}^{2} \alpha' \frac{\delta S_{\rm eff}}{\delta g_{ij}} = K_{ij}^{kl}  \beta_{kl} \,,
   }
where $K_{ij}^{kl}$ can be computed
perturbatively.  In the bosonic case, this was done in the minimal subtraction scheme in
\cite{Tseytlin:1986ti,Jack:1988rq}; in the heterotic case, this was done in a different scheme in \cite{FMRone,FMRtwo}; but we do not know of a minimal subtraction calculation of $K_{ij}^{kl}$ in the heterotic case.  In the bosonic case, $K_{ij}^{kl}$ receives
contributions starting at two loops, and it can be shown that this
is compatible with an independent calculation of the effective
action using scattering amplitudes.  The same thing may happen in the heterotic case: in particular, $R^3$ terms could indeed be absent from $S_{\rm eff}$, and the $g^4$ term in \eno{betafunction} could come entirely from $K_{ij}^{kl}$.  A similar conclusion is reached in \cite{efsOne,efsTwo}
where it is shown that the beta function of the heterotic string
in the presence of background gauge fields has a term at three
loops that behaves as $F^{3}$, even though no corresponding term is present in the the effective action.

Finally, it is possible to make a statement about the three-loop
structure of the beta function  in the $\alpha'$ expansion.
Excluding the Ricci tensor and the Ricci scalar, since the beta
function is computed within the minimal subtraction scheme, the
terms that are third order in the Riemann tensor and are compatible
with the $g^{4}$ terms in
 \eno{betafunction} are given by
 \eqn{Compatible}{
\alpha'^3 \left(\frac{1}{8} R_{klmn} R_i{}^{mlr} R_{j}{}^k{}_{lr} -\frac{1}{16} R_{i
kl j} R^{kmnr} R^l {}_{mnr} \right) \,.
  }

\subsection{Singularities of the critical exponents; central charge of the CFT}
\label{SINGULARITIES}

Because $\lambda$ involves products of $\Gamma$ functions it is
 natural to investigate the location of its singularities closest to the origin, as in
 \cite{fg}.  Because $\lambda_{1}^{het} = \frac{1}{2}
 \lambda_{1}^{bos}$, the location of the pole of $\lambda_{1}$
 coincides with the pole in the bosonic case, with half the
 residue.  One also has to note that $\eta_{1}(\mu)$ behaves as $\eta_{1} \sim
-\frac{4}{\pi^{2}}\frac{1}{2\mu-1}$ i.e. it has a simple pole at
$\epsilon =-1$.  But $\lambda_{1}$'s first singularity is at
$\epsilon=-3$, since the pole of $\eta_{1}$ is canceled by a similar
pole of $\chi_{1}$.  Examining term by term the structure of
$\lambda_{2}$ it is easy to see that the singularities of
$\lambda_{2}$ come from the $\eta_{1}^{2}$ factor that multiplies
the whole expression \eno{lambdatwo} and from the three-loop diagrams
that have the lagrange multiplier field propagator corrected, i.e.~$
\Pi_{2 \sigma}$, $\Pi_{3u}$, $F_{2u}$, and $F_{2\sigma}$.  Since
 \eqn{Rsingularities}{
 R_{3}(\mu) \sim \frac{-1}{2\mu-1},\qquad R_{2}(\mu) \sim
 \frac{1}{(2\mu-1)^{2}}
  }
 and $\lambda_{2}$ has terms that behave as $ \sim
R_{3}^{2}\eta_{1}^{2}$ and $R_{2}\eta_{1}^{2} $ times a $\mu$ polynomial
with no zero at $\mu=1/2$, we see that it has a fourth order pole.
In all, one finds
\eqn{lamdaOneSingular}{
 \lambda_{1} =- \frac{3/(4\pi ^{2})}{\epsilon+3} +\mathcal{O}(1)
 \quad \quad b_1\ = - \frac{\log(3+\kappa)}{2\pi^{2}}
 +\mathcal{O}(1)
 }
  \eqn{lamdaTwoSingular}{
 \lambda_{2} = \frac{8/\pi^{4}}{(\epsilon+1)^{4}} +
 \mathcal{O}((\epsilon+1)^{-3}) \quad \quad b_2\ = -
 \frac{16/3\pi^{4}}{(\kappa+1)^{3}} + \mathcal{O}((\kappa+1)^{-2})
  \,.
 }
The $1/(\kappa+1)^2$ term in $b_2$ comes only from the
factors $R_{3}\eta_{1}^{2}$ and from the Hartree-Fock
diagrams.  The singularities in the heterotic case are at the same locations and of the same order as in the bosonic case.

Because of the sign of $b_2$, there is clearly a zero of $\beta(g)$ (computed through order $1/D^2$) for negative $g$, close to $\kappa=-1$.  The same caveats discussed in \cite{fg} apply: higher order terms in $1/D$ could conceivably cause this zero to disappear or move significantly.  In section~\ref{DISCUSSION} we will comment further on higher-order corrections.  For the remainder of this section we will assume that the computation of $\beta(g)$ that we have carried out is precise enough to describe the zero correctly.

The zero of $\beta(g)$ arises through competition between the one-loop term (corresponding to Einstein gravity) and $b_2$ (corresponding to a combination of all $\alpha'$ corrections to Einstein gravity).  Because the geometry has string scale curvatures (more precisely, $L^2 \sim D\alpha'$) there is no reason to think that the worldsheet central charges are particularly close to the flat-space results.  Fortunately, one can calculate the central charges using Zamolodchikov's c-theorem:
  \eqn{ccharge}{
  \frac{\partial c}{\partial g} =  \frac{3(D+1)}{2g^{2}} \beta(g) \,.
  }
The result \eno{ccharge} holds for both the holomorphic and the anti-holomorphic sides: $c$ and
$\tilde{c}$ differ by a constant.  To derive the prefactor on the right hand side of \eno{ccharge},
one can consider two-point functions of the graviton perturbation
 $O_{ij} = \frac{1}{2\pi \alpha'}
 \partial X_i \bar{\partial} X_{j}+ \frac{1}{4 \pi} \Psi\partial \Psi $ around flat space, as is
 done in \cite{fg}.\footnote{Another way to derive \eno{ccharge} is to use the relation of the central charge to the spacetime effective action.  At least up to two-loop order, the effective action at the fixed point is equal to $-c/2\kappa_D^2 \alpha'$ \cite{Osborn:1988hd,Osborn:1989bu,deAlwis:1986rw}.  Also up to two loops, the $K^{kl}_{ij}$ of \eno{Seffts} is simply given by a product of Kronecker $\delta$'s,
  as in the bosonic case \cite{Jack:1988rq}.  Using the fact that in symmetric spaces
  \eqn{SymmetricSpaces}{
  \beta(g) g_{ij} = -g \beta_{ij} \qquad
  g^{ij}\frac{\partial}{\partial g^{ij}} = g
  \frac{\partial}{\partial g} \,,
  }
one indeed ends up with \eno{ccharge}.}
This prefactor receives higher loop corrections,
 and knowing $K^{kl}_{ij}$ in higher loops, one can in principal compute them.
   As in the bosonic and supersymmetric cases, the results suggest that
   with increasing $D$ the critical point moves
   closer to $\kappa =-1$: integrating \eno{ccharge} leads to
 \eqn{CentralChargeIntegral}{
 c &=(D+1) + \frac{3(D+1)}{2} \int_{0}^{\kappa_{c}}d \kappa \frac{1}{\kappa}
 \left(-\kappa +\frac{b_{1}(\kappa)}{D} +\frac{b_{2}(\kappa)}{D^{2}}\right)
   \approx   (D+1)(1-\frac{3}{2}\kappa_{c}) \cr
   \tilde{c} &= \frac{3}{2}(D+1) + \frac{3(D+1)}{2} \int_{0}^{\kappa_{c}}d \kappa \frac{1}{\kappa}
 \left(-\kappa +\frac{b_{1}(\kappa)}{D} +\frac{b_{2}(\kappa)}{D^{2}}\right) \approx \frac{3}{2}(D+1)(1-\kappa_{c}) \,,
 }
where we have noted that the central charge of the holomorphic side in flat space is $ c = D+1$, while for the anti-holomorphic side it is $ \tilde{c} =
\frac{3}{2}(D+1)$.  The approximate equalities arise from dropping the $b_1(\kappa)$ and $b_2(\kappa)$
terms from the integrand: their only role at this level of approximation is to set $\kappa_c$.
As $\kappa_c$ gets closer to $-1$ (i.e.~as $D$ becomes large), the central charges converge to
 \eqn{CentralChargeSat}{
 c = \frac{5}{2} (D+1) \qquad \tilde{c} = 3(D+1) \,.
 }
The result \eno{CentralChargeSat} for $c$ is the same as in the bosonic case, while for $\tilde{c}$ it
is the same as the type~II case \cite{fg}.  As in \cite{fg}, \eno{CentralChargeSat} appears to set only an
approximate upper bound on the central charges.  The dominant error in
the calculation~\ref{CentralChargeIntegral} is from the uncertainty in the prefactor in \eno{ccharge}.
Analogous to the speculations in \cite{fg}, it is conceivable that the expressions \eno{CentralChargeSat}
might in fact be exact.  But this would require a significant conspiracy between the prefactor in \eno{ccharge}
and the beta function.

   The
fact that the location of the critical point at finite $D$ is so close
to the singularity of $\lambda$ means that the critical exponent
$\lambda$ evaluated at the critical point is large and positive.
This leads to an operator with a large and negative dimension, which
appears to violate unitarity. However, one could hope that a
consistent GSO projection would project this
operator out of the spectrum.

\section{Discussion}
\label{DISCUSSION}

The existence of the $AdS_{D+1}$ critical point depends on competition between one-loop and $1/D^2$ effects.  It would therefore be instructive to compute the beta function through order $1/D^3$ and see whether the fixed point persists.  Given that the number of diagrams needed for the
  computation at the next order grows significantly, the shortest
  path seems to be calculating $\chi_{3}$ and using \eno{scalingma}
  to deduce $\lambda_{3}$. However, note that for the calculation of
  $\chi_{3}$ one needs to derive the residues of diagrams at order $1/D^4$, which include some six-loop diagrams.

There is some reason to think that the singularities of $\lambda$ at order $1/D^3$ are no worse than at order $1/D^2$: examining the diagrams needed for
  the Dyson equations of the Lagrange multiplier fields, we see that
  at order $1/D^3$ these come from either inserting a $\sigma$ or $u$
  propagator in the $1/D^{2}$ diagrams or inserting a loop of
  $S$ or $\Psi$ in the middle of the diagram. The computation for the
  diagrams that come from inserting a $\sigma$ or $u$ propagator can
  easily be seen to be reduced to the sum of diagrams similar to
  $\Pi_{2}$ or $\Pi_{3}$ with one different exponent.  A naive calculation
  does not produce any worse singularities than the ones already
  contained in  $\Pi_{2}$ and $\Pi_{3}$.  However, one also has to compute
  the more difficult diagrams with the additional $S$ or $\Psi$ loop.

  It is evident that the methods of \cite{Vone,Vtwo}, has many
advantages over calculating Feynman diagrams in momentum space.  In
the latter approach one encounters difficulties already in calculating
second order diagrams, since the propagators of the Lagrange
multiplier fields are in general hypergeometric functions.  It is
noteworthy that even though we start from $ d =2+\epsilon$
dimensions, one can calculate critical exponents of the $O(N)$ model
in any dimension $ 2 <d< 4 $, and there is agreement with the
results in three dimensions \cite{Okabe:1978mp} in the bosonic case.

Perhaps the methods of \cite{Vone,Vtwo} could be applied to a related quantum field theory:
 \eqn{KPModel}{
 S = \int d^{d} x \, \left( \frac{1}{2} \nabla  \vec{\Phi} \cdot
 \nabla\vec{\Phi} + \frac{1}{2} \lambda \sigma \vec{\Phi} \cdot \vec{\Phi}
 - \frac{\lambda N}{4} \sigma ^{2} \right) \,,
 }
which for $d=3$ is the proposed dual of an $AdS_4$ vacuum of a theory with arbitrarily high spin gauge fields \cite{kpON}.  What makes \eno{KPModel} susceptible to a position-space treatment analogous to those in \cite{Vone,Vtwo} is that only cubic vertices are involved.  It would be interesting to compute, for example, the four point function of $\sigma$ to order $1/N^{2}$ and compare it to the corresponding $AdS_4$ calculation, as
is done for example in \cite{LMR} at order $1/N$.

\section*{Acknowledgements}

GM would like to thank J.~Friess for valuable conversations.

\appendix

\section{Anomalies}
\label{ANOMALIES}

Since we have coupled only the right moving fermions to gravity it
is natural to investigate whether there are anomalies.  These are
related to a breakdown of general coordinate invariance or local Lorentz
invariance.  We will investigate only the latter, as is
usually done \cite{Greenone}.  Indeed when there is a coordinate
anomaly one can add a counterterm to the action and convert it to a
Lorentz anomaly \cite{Bardeen}.  It is convenient to use the tetrad
formalism.  Local Lorentz transformations in this formalism are
given by
 \eqn{tetra}{
  e' _{\mu}{}^{p}(x) =
e_{\mu}{}^{q}(x)\Theta^{p}{}_{q}(x) \,.
 }
 The Riemann tensor can be written
$R_{\mu \nu} {}^{p} {}_{q}$, with mixed spacetime and tangent space
indices, and can be regarded as a two form $R_{2}$. We only have to
worry about the massless fields of the supergravity sector
\cite{Greenone}. The anomaly polynomials for the spinor and the
gravitino contain only terms that are proportional to polynomials in $\Tr R_2^{\wedge 2m}$.
  In the AdS space that we are interested in we can calculate
 \eqn{wedgeAdS}{
R_{\mu \nu} {}^{a}{}_ {b} R_{\kappa \lambda}{} ^{c}{}_ {a} =
\frac{1}{L^{2}} ( \delta^{c}{}_{k} R_{\mu\nu\lambda b} +
\delta^{c}{}_{\lambda} R_{\mu\nu b \kappa})
 }
 which when antisymmetrizing to get the wedge product
returns zero. So $ R_{2} \wedge R_{2} =0$ in our case, and we do not
have to worry about the gravitational anomalies.  Another way to
view this is to say that the field strength $H_3 = d B_2$ obeys the
modified Bianchi identity
 \eqn{modifiedbianchi}{
  dH_3 = \frac{1}{4\pi}( \Tr R _{2} \wedge
R_{2}-\Tr F_{2}\wedge F_{2}) \,.
 }
 A three form $H_3$ obeying \eno{modifiedbianchi} is required for
cancelation of perturbative heterotic string worldsheet anomalies,
as is briefly reviewed in  \cite{Wittenanom}.  Because $\Tr R_{2}
\wedge R_{2} =0 $, this is trivially satisfied.

\section{Position space methods for calculating graphs}\label{method}
\label{METHODS}

\begin{figure}[t]
\centering
\includegraphics[scale=0.9]{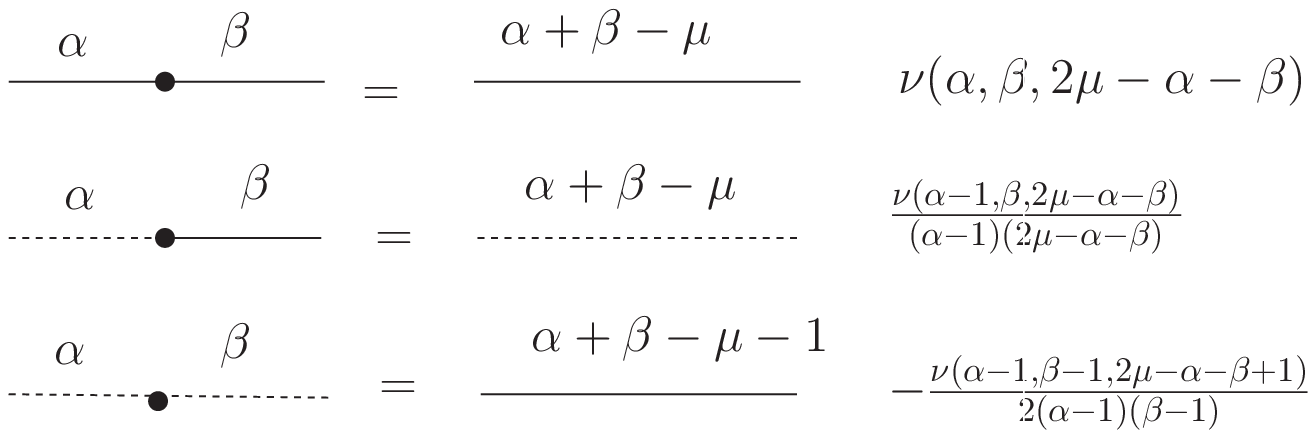}
\caption{Products of two propagators are related to a single
propagator.  So, in a two-loop diagram,($\Sigma_{1}$ for example)
inserting a point in one of the three propagators that connect to
one internal vertex can make this vertex unique.  The dotted lines
denote fermions.  Since we are dealing with chiral fermions, taking
the trace in the third graph only produces one half the full result.
$\nu$ is equal to $\nu(x_{1},x_{2},x_{3}) = \pi \prod _{i=1}^{3}
\alpha (x_{i})$.}\label{chains}
\end{figure}

\begin{figure}[t]
\centering
\includegraphics[scale=1.0]{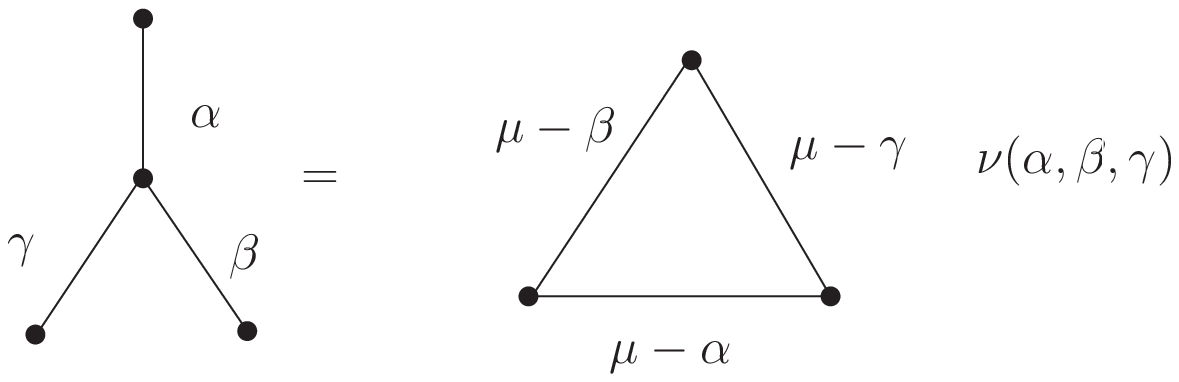}
\caption{An identity that allows the integration of a unique vertex.
Only bosonic propagators are shown. Similar identities with fermions
can which are used in our calculation can be found in
\cite{graceyFour}}\label{triangles}
\end{figure}
There are 11 diagrams needed for the calculation of $\eta$ at
$1/D^{2}$. We designate them by $\Sigma$, $\Pi$, $\Phi$, $F$. The way to
compute them was developed in \cite{Vone,Vtwo} for the bosonic
graphs and extended to include fermionic graphs in
\cite{graceyOne,graceyTwo}. The main advantage of the method is that
there is no need to explicitly evaluate any Feynman diagram.  Here
we will only give a few key observations that facilitate the
evaluation of the diagrams. The first observation is that the chain
of two propagators is equal to a propagator times a prefactor.
Graphically this is shown in Fig \ref{chains}, where
$\nu(x_{1},x_{2},x_{3}) = \pi^{\mu} \prod_{i=1}^{3} \alpha(x_{i})$.
The third exponent $x_{3}$ is determined by the ``uniqueness''
\cite{Vtwo} requirement $\sum_{i}x_{i} =2\mu$, for the bosonic
graphs and $ \sum_{i} x_{i} = 2\mu -1$ if there are one or more
fermion lines in the graph.  An identity exists for a three point
vertex, which is similarly related to a ``unique'' triangle, where now
the uniqueness requirement is that $ \sum_{i} x_{i} = \mu $.  If
there are one or more fermion lines the uniqueness changes to $
\sum_{i}x_{i} = \mu+1$, and the results of \cite{graceyOne} are
unchanged in our case.  There is no similar identity for a four point
function, and for the (1,1) supersymmetric model that means that one
has to retain the auxiliary field $F$.  In computing the values at
order $1/D^{2}$, $ \chi$ is set to zero \cite{Vtwo}. For a non-zero
$\Delta$ the diagrams, for example the self energy of $\Psi$
designated $A$, lose their uniqueness.  However one can subtract from
$A$ a graph $B$ that has the same divergent substructure as $A$, but can
be calculated for an arbitrary $\Delta$. So one has to compute $(A-B)
+B$.  Since $B$ can be calculated for arbitrary $\Delta$ and contains
the divergence, one can evaluate $(A-B)$ at zero $\Delta$ when both
diagrams become unique. A valuable first step in the calculation is
the evaluation of all the self-energy graphs,  which we will not
include here since it was done in detail in \cite{Vtwo,graceyTwo}.
 We just note that the most basic tool is the insertion of a point
facilitated by the fact that we can write a propagator as a product
of two.  Then one can choose one of the exponents in such a way that
the vertex that the propagator is attached to, becomes unique.

\section{Calculation of the graphs needed for $\eta_2$}
\label{ETADETAILS}

We give the results for the various graphs occurring in the
$1/D^{2}$ calculation. We only give the simple pole term and the
constant term in an expansion in $\Delta$. The purely bosonic graphs
were calculated in \cite{Vtwo}.  Compared to the calculation
of the fermionic graphs in \cite{graceyFour}, there are differences
that have to do with taking the trace of fermion loops, i.e.~some
factors of two in bosonic diagrams with fermion loops.  Otherwise the
calculation is almost identical.  We find small discrepancies with \cite{graceyFour} in some
of the diagrams, mostly factors of $2$ and some minus signs. The
most notable difference is $\Phi_{1}$, where the residue has a
different denominator. We believe that our value is correct, since
it leads to the same $\chi_{1}$ as the evaluations from the other
equations \eno{mazaSeveral}.
  \eqn{betadef}{
 B(x) = \psi(x) + \psi(\mu-x)
 }
  \eqn{sigmaone}{
 \Sigma_{1} = \frac{ 2\pi^{2\mu} \alpha^{2}
(\Delta_{S}) \alpha(\Delta_{\sigma}) }{ \Delta \Gamma(\mu)} \left( 1
+\frac{\Delta}{2}\left[B(\Delta_{\sigma})-B(\Delta_{S})\right]\right)
 }
  \eqn{sigmatwo}{
 \Sigma_{2}  = \frac{ 2 \pi^{2\mu}
\alpha^{2}(\Delta_{S}) \alpha(\Delta_{\sigma}-1)}{ \Delta
\alpha(\Delta_{\sigma}-1) \Gamma(\mu)}\left( 1+ \frac{\Delta}{2} \left[
B(\Delta_{\sigma}-1)-B(\Delta_{S}) + \frac{1}{\Delta_{\sigma}-1} -
\frac{1}{\Delta_{S}}\right] \right)
 }
  \eqn{sigmathree}{
 \Sigma_{3} = \frac{ 2\pi^{4\mu} \alpha^{3}(\Delta_{S})
\alpha^{3}(\Delta_{\sigma})\alpha(\mu+\Delta_{S}-\Delta_{\sigma})}{\Delta
\Gamma(\mu)} \left( \frac{1}{2} + \Delta \left[
B(\Delta_{\sigma})-B(\Delta_{S})\right]\right)
}
  \eqn{sigmafour}{
  & \Sigma_{4} = \frac{\pi^{4\mu} \alpha^{3}
 (\Delta_{S}) \alpha^{2}(\Delta_{\sigma}-1) \alpha(\Delta_{\sigma})
 \alpha(\Delta_{S}+\mu -\Delta_{\sigma})}{\Delta
 \Delta_{S}(\Delta_{S}+\mu-\Delta_{\sigma})(\Delta_{\sigma}-1)^{2}
 \Gamma(\mu)} \cr & \times{}\left(1 + \frac{\Delta}{2}\left[
 B(\Delta_{\sigma})+ 3 B(\Delta_{\sigma}-1)- 4 B(\Delta_{S}) +
 \frac{3}{\Delta_{\sigma}-1} - \frac{2}{\Delta_{S}}\right] \right)
  }
  \eqn{Pione}{
   \Pi_{1} =\frac{2 \pi^{2\mu} \alpha^{2}(\Delta_{S})
 \alpha(\Delta_{\sigma})}{\Delta \Gamma(\mu)} \left( 1 + \Delta \left[
 B(\Delta_{\sigma}) -B(\Delta_{S}) \right] \right)
 }
  \eqn{Pitwo}{
  \Pi_{2} = \frac{ \pi^{4\mu} \alpha^{3}(\Delta_{S})
 \alpha^{3}(\Delta_{\sigma}) \alpha(\Delta_{S}+\mu
 -\Delta_{\sigma})}{ \Delta \Gamma(\mu)} \left( 1+ \Delta\left[
 4B(\Delta_{\sigma}) -3B(\Delta_{S}) -
 B(\mu+\Delta_{S}-\Delta_{\sigma})\right]\right)
  }
  \eqn{Pithree}{
  \Pi_{3} & = \frac{ \pi^{4\mu} \alpha^{3}(\Delta_{S})
 \alpha^{2}(\Delta_{\sigma}-1) \alpha(\Delta_{S}+\mu
 -\Delta_{\sigma})}{ 2\Delta \alpha(\mu-\Delta_{\sigma})
 \Delta_{S}(\Delta_{S}+\mu-\Delta_{\sigma})(\Delta_{\sigma}-1)^{2}
 \Gamma(\mu)}  \cr &\qquad{} \times
 \bigg( 1 + \Delta \Big[ 2B(\Delta_{\sigma}) -  3
 B(\Delta_{S}) + 2B(\Delta_{\sigma}-1)
 -B(\Delta_{S}+\mu-\Delta_{\sigma})  \cr &\qquad{} - \frac{1}{\Delta_{S}} +
 \frac{2}{\Delta_{\sigma}-1} -
 \frac{1}{\Delta_{S}-\Delta_{\sigma}+\mu}\Big]  \bigg)
 }
 \eqn{Phione}{
 \Phi_{1} = - \frac{ \pi^{2\mu} \alpha^{2} (\Delta_{S}-1)
 \alpha(\Delta_{\sigma})}{\Delta \Delta_{S}(\Delta_{S}-1)
 \Gamma(\mu)} \left( 1+ \Delta\left[ B(\Delta_{\sigma})-B(\Delta_{S}-1)
 - \frac{1}{\Delta_{S}-1}\right] \right)
 }
  \eqn{Phitwo}{
 \Phi_{2} = - \frac{ \pi^{4\mu} \alpha^{3}(\Delta_{S})
 \alpha^{2}(\Delta_{\sigma}-1)
 \alpha(\Delta_{\sigma})\alpha(\Delta_{S}+\mu-\Delta_{\sigma})}{\Delta
 \Delta_{S}(\Delta_{S}+\mu-\Delta_{\sigma})(\Delta_{\sigma}-1)^{2}
 \Gamma(\mu)}  \cr
 \times {} \left( 1+ \Delta \left[ B(\Delta_{\sigma})
 -2B(\Delta_{S}) + B(\Delta_{\sigma}-1)
 +\frac{1}{\Delta_{\sigma}-1}\right]\right)
 }
  \eqn{Fone}{
 F_{1} = -\frac{2\pi^{2\mu} \alpha^{2}(\Delta_{S})
\alpha(\Delta_{\sigma}-1)}{\Delta \Delta_{S} (\Delta_{\sigma}-1)
\Gamma(\mu)}\left( 1+ \Delta\left[ B(\Delta_{\sigma}-1)- B(\Delta_{S})
- \frac{1}{2\Delta_{S}} + \frac{1}{\Delta_{\sigma}-1}\right]\right)
 }
  \eqn{Ftwo}{
& \qquad  \qquad F_{2} = - \frac{ \pi^{4\mu} \alpha^{3} (\Delta_{S})
\alpha(\Delta_{\sigma}) \alpha^{2}(\Delta_{\sigma}-1)
\alpha(\Delta_{S}+\mu-\Delta_{\sigma})}{ \Delta \Delta_{S}
(\Delta_{S}+\mu -\Delta_{\sigma})(\Delta_{\sigma}-1)^{2}
\Gamma(\mu)}  \cr
 & \times{}\bigg( 1 + \Delta\bigg[ B(\Delta_{\sigma})+3 B(\Delta_{\sigma}-1) -
 B(\Delta_{S}
+\mu-\Delta_{\sigma}) -3 B(\Delta_{S}) \cr
&\qquad \quad - \frac{1}{\Delta_{S}} +
\frac{3}{\Delta_{\sigma}-1} - \frac{1}{\Delta_{S} + \mu
-\Delta_{\sigma}} \bigg]  \bigg) \,.
}
Note that there is a similar three-loop diagram with $\Pi_{3}$ with
the role of the $\Psi$, $u$ propagators interchanged.  But it scales as
$1/N^{3}$.  A very useful identity for the evaluation of various
quantities given in the text is $B(x) =B(\mu-x)$.

\section{Calculation of the graphs need for $\lambda_2$}
\label{LAMBDADETAILS}

In this section we give the formal expressions for the sums of the
corrected diagrams \eno{SigmaS}-\eno{FuFsigma}, the values of the
43 individual diagrams that contribute \eno{SigmaoneSab}-\eno{Ftwou}, and finally the explicit form of the sums
\eno{SigmaSsum}-\eno{Fusum}.  Beforehand one has to define the
functions appearing as
 \eqn{Rdefinition}{
  &R_{1}= \psi'(\mu-1)-\psi'(\mu)
  \cr
 &R_{2} = \psi'(2\mu-3)-\psi'(2-\mu)-\psi'(\mu-1)+\psi'(1) \cr
 &R_{3} = \psi(2\mu-3) + \psi(2-\mu) -\psi(\mu-1)
-\psi(1) \,,
 }
where $\psi(z) = d \log \Gamma(z) / dz$.
  \eqn{SigmaS}{
 \Sigma_{S} = 2 w^{2} \Sigma_{1Sa} + w^{2} \Sigma_{1Sb}
- v^{2} \Sigma_{2S} +&N w^{3}
(2\Sigma_{3Sa}+2\Sigma_{3Sb}+\Sigma_{3Sc}) \cr-
& 2Nv^{2}w(\Sigma_{4Sa}+\Sigma_{4Sb} +\Sigma_{4Sc})
 }
  \eqn{SigmaPsi}{
\Sigma_{\Psi} = - 2v^{2} \Sigma_{2\Psi} -2N v^{2}
w(\Sigma_{4\Psi a } +\Sigma_{4\Psi b})
 }
\eqn{Sigmasigma}{ \Sigma_{\sigma} = 2w^{2} \Sigma_{1\sigma} + N
w^{3}(2 \Sigma_{3\sigma a} + \Sigma_{3\sigma b}) - 2N v^{2}w
\Sigma_{4\sigma}
 }
\eqn{Sigmau}{ \Sigma_{u} = -2 v^{2} \Sigma_{2u} - 2N
v^{2}w(\Sigma_{4ua} +\Sigma_{4ub})
 }
\eqn{Phipsi}{ \Phi_{\Psi} = v^{2} \Phi_{1\Psi} + N v^{2}w
\Phi_{2\Psi}\quad \quad \Phi_{\sigma} = Nv^{2}w\Phi_{2\sigma}
 }
\eqn{PhiSPhiu}{
 \Phi_{S} = 2 v^{2}\Phi_{1S} + 2Nv^{2}w(
\Phi_{1Sa}+\Phi_{2Sb}) \quad\quad  \Phi_{u} = 2 v^{2} \Phi_{1u} + N
v^{2}w \Phi_{2u}
 }
\eqn{PiS}{
 \Pi_{S} = 4 w^{2} \Pi_{1S} + Nw^{3}( 4
\Pi_{2Sa}+2\Pi_{2Sb}) -4 Nwv^{2}\Pi_{3S}
 }
\eqn{PisigmaPiu}{
 \Pi_{\sigma} = w^{2} \Pi_{1\sigma} + 2 Nw^{3} \Pi_{2
\sigma}\quad \quad \Pi_{\Psi} = -2 N wv^{2} \Pi_{3\Psi}\quad\quad
\Pi_{u} = -2 Nwv^{2} \Pi_{3 u}
 }
\eqn{FpsiFS}{
 F_{\Psi} = 2 v^{2} F_{1\Psi} +  2N v^{2}w 2F_{2\Psi
}\quad \quad F_{S}= 2 v^{2} F_{1S} + 2Nv^{2}w ( F_{2Sa} + F_{2Sb})
 }
\eqn{FuFsigma}{
 F_{u} = v^{2} F_{1u} + Nv^{2}w F_{2u}\quad \quad F_{\sigma} =
Nv^{2}wF_{2\sigma}
 }
There are 19 diagrams associated with the S-propagator and eight for
the propagator of the other fields.  The value of each graph of each
graph is  given below for completeness.  The bosonic ones were
calculated in \cite{Vtwo}, while similar to the fermionic ones were
done in \cite{graceyThree}.
 \eqn{SigmaoneSab}{
 \Sigma_{1Sa} = \frac{\pi^{2\mu}}{ (\mu-2)
\Gamma^{2}(\mu)} \quad \quad \Sigma_{1Sb} = \frac{ \pi^{2\mu}}{
(\mu-2)^{2} \Gamma^{2}(\mu)}
 }
  \eqn{SigmatwoS}{ \Sigma_{2S} =-
\frac{2(\mu+1)\pi^{2\mu}}{\mu(\mu-1)\Gamma^{2}(\mu)}\quad \quad
\Sigma_{3Sa} = \frac{ ( \mu^{2} -3\mu +1) \Gamma(1-\mu)
\pi^{4\mu}}{(\mu-2)^{3} \Gamma(\mu) \Gamma(2\mu-3)}
  }
  \eqn{Sigmathreeb}{ \Sigma_{3Sb} = \frac{\pi^{4\mu}
\Gamma(2-\mu)}{(2-\mu)\Gamma(\mu-1)\Gamma(2\mu-2)}\left( 3R_{1} +
\frac{2\mu-3}{(\mu-2)^{2}}\right)
  }
  \eqn{Sigmathreec}{
 \Sigma_{3Sc} = \frac{ \pi^{4\mu}
\Gamma(4-\mu)}{(\mu-2)^{3} \Gamma(\mu-1)\Gamma(2\mu-4)}\quad \quad
\Sigma_{4Sa} = \frac{ 2\pi^{4\mu} \Gamma(1-\mu)}{\Gamma(\mu)
\Gamma(2\mu-2)}
  }
  \eqn{Sigmafourb}{
\Sigma_{4Sb} = - \frac{ \pi^{4\mu}
\Gamma(2-\mu)}{\Gamma(\mu)\Gamma(2\mu-2)} \left( 3R_{1} +\frac{
2\mu-3}{(\mu-1)(\mu-2)}\right)
  }
  \eqn{Sigmafourc}{
 \Sigma_{4Sc} = \frac{2 (\mu-3)(2\mu-3)
\Gamma(1-\mu) \pi^{4\mu}}{(2-\mu)\Gamma(2\mu-2)\Gamma(\mu)}\quad
\quad \Sigma_{2\Psi} = \frac{ \pi^{2\mu} }{(\mu-1)\Gamma^{2}(\mu)} =
- \Phi_{2S}
 }
  \eqn{SigmafourPsia}{
 \Sigma_{4\Psi a} = \frac{ \pi^{4\mu}
\Gamma(1-\mu) ( 2\mu^{2} -5\mu
+1)}{2(\mu-2)\Gamma(\mu)\Gamma(2\mu-2)} = - \Phi_{2Sa} \quad \quad
\Sigma_{4\Psi b} = \frac{ \pi^{4\mu} 3( \mu-3)\Gamma(2-\mu)
R_{1}}{2(2-\mu) \Gamma(\mu)\Gamma(2\mu-2)} = - \Phi_{2Sb}
 }
  \eqn{Sigmaonesigma}{
\Sigma_{1\sigma} = \frac{\pi^{2\mu} ( \mu^{2}
-3\mu +1)}{(\mu-2)^{2}\Gamma^{2}(\mu)} = \Pi_{1S}\quad \quad
\Sigma_{3\sigma a} = \frac{\pi^{4\mu}( 2\mu^{2} -7\mu+4)
\Gamma(1-\mu)}{(\mu-2)^{3}\Gamma(\mu)\Gamma(2\mu-3)} =  \Pi_{2Sa}
 }
  \eqn{Sigmathreesigma}{
 \Sigma_{3\sigma b} = \frac{ 3
\pi^{4\mu}\Gamma(3-\mu)
R_{1}}{(2-\mu)^{3}\Gamma(\mu-1)\Gamma(2\mu-2)} =  \Pi_{2Sb}
\quad\quad \Sigma_{4\sigma } = \frac{  \pi^{4\mu}
(2\mu-5)\Gamma(1-\mu)}{(\mu-2)\Gamma(\mu)\Gamma(2\mu-2)} = \Pi_{3S}
 }
  \eqn{Sigmatwou}{
  \Sigma_{2u} =
\frac{\pi^{2\mu}}{\Gamma^{2}(\mu)}\quad\quad \Sigma_{4ua} =
\frac{\pi^{4\mu}
(4\mu-9)\Gamma(1-\mu)}{2(\mu-2)\Gamma(\mu)\Gamma(2\mu-2)} = -
F_{2sa}
  }
  \eqn{Sigmafouru}{
\Sigma_{4ub } = \frac{ 3 \pi^{4\mu}
\Gamma(2-\mu)R_{1}}{2(\mu-2)\Gamma(\mu) \Gamma(2\mu-2)} = - F_{2Sb}
 }
  \eqn{PhioneS}{
\Phi_{1S} = - \frac{
\pi^{2\mu}}{(\mu-1)\Gamma^{2}(\mu)}\quad\quad \Phi_{1u} =
\frac{\pi^{2\mu} \mu}{(1-\mu)\Gamma^{2}(\mu)}
 }
  \eqn{Phitwou}{
\Phi_{2u} = \frac{\pi^{4\mu} \Gamma(1-\mu) (
4\mu^{2}-11m+5)}{2(1-\mu)(\mu-2)\Gamma(\mu)\Gamma(2\mu-2)} \quad
\quad \Pi_{1\sigma} = \frac{3
\pi^{2\mu}R_{1}}{(2-\mu)(2\mu-3)\Gamma^{2}(\mu-1)}
 }
  \eqn{Pitwosigma}{
 \Pi_{2\sigma} = \frac{ \pi^{4\mu} \Gamma(2-\mu)}{
2 (2-\mu)^{3} \Gamma(\mu-1)\Gamma(2\mu-1)} \left( 6R_{1} -R_{2}
-R_{3}^{2} + \frac{2(\mu-2)}{(\mu-1)(3\mu-2)}(R_{3}
-\frac{1}{\mu-2})\right)
 }
  \eqn{Pithreeu}{
 \Pi_{3u} = -\frac{
\Gamma(2-\mu)\pi^{4\mu}}{4(\mu-2)^{2}\Gamma(\mu)\Gamma(2\mu-2)}
\left(6R_{1}^{2}-R_{2}-R_{3}^{2}+\frac{2R_{3}(\mu-2)-2}{(\mu-1)(2\mu-3)}\right)
=- F_{2\sigma}
 }
  \eqn{PithreePsi}{
 \Pi_{3\Psi} = \frac{3\Gamma(2-\mu)\pi^{4\mu}
R_{1}}{(2\mu-3)\Gamma(\mu)\Gamma(2\mu-1)} = \Phi_{2\sigma} \quad
\quad F_{1\Psi} = \frac{\mu \pi^{2\mu}}{ (1-\mu)\Gamma^{2}(\mu)}
 }
 \eqn{FtwoPsi}{
 F_{2\Psi } = \frac{ \pi^{4\mu}(4\mu^{2}-11 \mu
+5)\Gamma(1-\mu)}{2 (1-\mu)(\mu-2)\Gamma(\mu)\Gamma(2\mu-2)}
  }
  \eqn{FoneS}{
 F_{1S} = -\frac{\pi^{2\mu}}{\Gamma^{2}(\mu)} \quad
\quad F_{1u}  = \frac{3(\mu-1) \pi^{2\mu}}{2(\mu-2) \Gamma^{2}(\mu)}
 }
 \eqn{Ftwou}{
 F_{2u} = \frac{\pi^{4\mu}
 \Gamma(2-\mu)}{4(\mu-1)^{2}(\mu-2)^{2}\Gamma(\mu)
 \Gamma(2\mu-2)}\bigg( (\mu-1)( 6R_{1}-R_{2}-R_{3}^{2})\cr  +
 \frac{2(\mu-2)R_{3}}{2\mu-3} +\frac{2}{2\mu-3} -
 \frac{4}{\mu-1}\bigg) \,.
 }

\subsection{Summing up graphs}
\label{SUMDETAILS}

In this subsection we give the explicit values for the various sums
appearing in the $\lambda_{2}$ calculation. We have omitted the
$1/N^2$ factor that multiplies all of the diagrams so that
the expression \eno{lambdatwo} for $\tilde{\lambda}_{2}$, does not
contain any factors of $N$.
  \eqn{SigmaSsum}{
 \frac{\Sigma_{S}}{\eta_{1}v_{1}} = -
1+\frac{\mu+2}{\mu-1} +&\frac{2}{(\mu-1)^2} +2 \mu(2\mu-5)
+2\mu(2\mu-3)(\mu-3)(2-(\mu-2)^{2}) \cr
 &-4
\mu(\mu-2)-6\mu(\mu-2)R_{1}
 }
  \eqn{SigmaPsisum}{
 \Sigma_{\Psi} =
-\eta_{1}v_{1}\left(\frac{\mu^{2}(\mu-2)(2\mu-3)}{\mu-1}+3(\mu-1)(\mu-3)R_{1}\right)
= \Phi_{S}
 }
 \eqn{Sigmasigmasum}{
 \Sigma_{\sigma} = -\eta_{1}v_{1} \left(
\frac{\mu(3+\mu(3\mu-7))}{(\mu-1)^{2}} + 3\mu(\mu-2)R_{1} \right)
=\frac{\Pi_{S}}{2}
 }
  \eqn{Sigmausum}{ \Sigma_{u} =
\eta_{1}v_{1}(\mu(8-4\mu+3(\mu-1)R_{1}) = F_{S}
 }
  \eqn{Phisigmasum}{
 \Phi_{\sigma }=
\eta_{1}v_{1}\frac{3\mu(\mu-1)(\mu-2)R_{1}}{2\mu-3} = (\Pi_{\Psi})/2
 }
  \eqn{Phiusum}{
 \frac{\Phi_{u}}{\eta_{1}v_{1}} =
1+6\mu-4\mu^{2} -\frac{1}{\mu-1} = - \frac{F_{\Psi}}{\eta_{1}v_{1}}
 }
  \eqn{Pisigmasum}{
 \frac{\Pi_{\sigma}}{\eta_{1}v_{1}} = \left(
\frac{3\mu(\mu-2)}{2(2\mu-3)} R_{1} -\frac{\mu}{\mu-1}\left[
6R_{1}-R_{2}-R_{3}^{2}+\frac{2((\mu-2)R_{3}- 1)}{(2\mu-3)(\mu-1)}
\right] \right)
  }
 \eqn{Piusum}{
 \frac{\Pi_{u}}{\eta_{1}v_{1}} =
\frac{\mu(\mu-1)}{\mu-2} \Big( 6R_{1} -R_{2}-R_{3}^{2} +
2\frac{(\mu-2)R_{3} -1}{(2\mu-3)(\mu-1)}\Big) = -\frac{2F_{\sigma}}{
\eta_{1}v_{1}}
 }
 \eqn{Fusum}{
 \frac{F_{u}}{\eta_{1}v_{1}} = \frac{3}{4}\frac{\mu(\mu-1)}{\mu-2}
+\frac{\mu}{4}\left(6R_{1}-R_{2}+R_{3}^{2}-\frac{4}{(\mu-1)^{2}}+\frac{2}{(2\mu-3)(\mu-1)}
+\frac{2(\mu-2)}{2\mu-3} R_{3}\right) \,.
  }

\bibliographystyle{ssg}
\bibliography{heterotic}
\end{document}